\documentclass[longauth]{aa} 

\usepackage[varg]{txfonts}
\usepackage{tabularx}
\usepackage{booktabs}

\usepackage{hyperref}
\usepackage{multirow}
\usepackage{subfig}
\usepackage{siunitx}
\usepackage{tikz}

\usepackage{graphicx,amsmath}
\usepackage{amssymb}
\usepackage{color}
\usepackage{natbib}		
\usepackage{url}
\usepackage{threeparttable}   
\usepackage{soul,mathabx}
\usepackage{enumitem}
\usepackage[version=4]{mhchem}

\def\approxsup{%
  \def\p{%
    \setbox0=\vbox{\hbox{$>$}}%
    \ht0=0.6ex \box0 }%
  \def\s{%
    \vbox{\hbox{$\sim$}}%
  }%
  \mathrel{\raisebox{0.7ex}{%
      \mbox{$\underset{\s}{\p}$}%
    }}%
}

\def\approxinf{%
  \def\p{%
    \setbox0=\vbox{\hbox{$<$}}%
    \ht0=0.6ex \box0 }%
  \def\s{%
    \vbox{\hbox{$\sim$}}%
  }%
  \mathrel{\raisebox{0.7ex}{%
      \mbox{$\underset{\s}{\p}$}%
    }}%
}

\bibpunct{(}{)}{;}{a}{}{,}           

\begin{document}

   \title{ATREIDES I} 
   \subtitle{Embarking on a trek across the exo-Neptunian landscape \\ with the TOI-421 system}

   \author{
   V. Bourrier\inst{\ref{inst:1}},
   M. Steiner\inst{\ref{inst:1}}, 
   A. Castro-Gonz\'alez\inst{\ref{inst:11}},
   D.~J. Armstrong\inst{\ref{inst:9},\ref{inst:10}},  
   M. Attia\inst{\ref{inst:1}},   
   S. Gill\inst{\ref{inst:9},\ref{inst:10}},             
   M. Timmermans\inst{\ref{inst:21},\ref{inst:20}},     
   J. Fernandez\inst{\ref{inst:9},\ref{inst:10}},     
   F. Hawthorn\inst{\ref{inst:9},\ref{inst:10}},          
   A.~H.~M.~J. Triaud\inst{\ref{inst:20}},              
   F. Murgas\inst{\ref{inst:22},\ref{inst:23}},   	    
   E. Palle\inst{\ref{inst:22},\ref{inst:23}},          
   H. Chakraborty\inst{\ref{inst:1}},                   
   K. Poppenhaeger\inst{\ref{inst:8},\ref{inst:12}},   
   M. Lendl\inst{\ref{inst:1}},                         
   D.~R. Anderson\inst{\ref{inst:9},\ref{inst:10},\ref{inst:42}},     
   E.~M. Bryant\inst{\ref{inst:9},\ref{inst:30}},        
   E. Friden\inst{\ref{inst:1}},   
   J.~V. Seidel\inst{\ref{inst:2},\ref{inst:3}},  
   M.~R. Zapatero Osorio\inst{\ref{inst:11}},               
   F. Eeles-Nolle\inst{\ref{inst:9},\ref{inst:10}},  
   M. Lafarga\inst{\ref{inst:9},\ref{inst:10}}, 
   I.~S. Lockley\inst{\ref{inst:9},\ref{inst:10}},  
   J. Serrano Bell\inst{\ref{inst:14}},       
   R. Allart\thanks{SNSF Postdoctoral Fellow}\inst{\ref{inst:27}},    
   A. Meech\inst{\ref{inst:18}},          
   A. Osborn\inst{\ref{inst:28}},	             
   R.~F. D\'iaz\inst{\ref{inst:25},\ref{inst:14}},     
   M. A. Fetzner Keniger\inst{\ref{inst:9},\ref{inst:10}},  
   G. Frame\inst{\ref{inst:9},\ref{inst:10}},  
   A. Heitzmann\inst{\ref{inst:1}},     
   A. Ringham\inst{\ref{inst:6}}, 
   P. Eggenberger\inst{\ref{inst:1}},    
   Y. Alibert\inst{\ref{inst:16},\ref{inst:15}},    
   J.~M. Almenara\inst{\ref{inst:1},\ref{inst:19}},  
   A. Leleu\inst{\ref{inst:1}},   
   S.~G. Sousa\inst{\ref{inst:4}},     
   S.~J. Mercier\inst{\ref{inst:24},\ref{inst:1}},       
   V. Adibekyan\inst{\ref{inst:4},\ref{inst:13}},   
   M.~P. Battley\inst{\ref{inst:6},\ref{inst:1}},   
   E. Delgado Mena\inst{\ref{inst:11},\ref{inst:4}},    
   W. Dethier\inst{\ref{inst:4}},      
   J.~A. Egger\inst{\ref{inst:15}},   
   K. Barkaoui\inst{\ref{inst:21},\ref{inst:24},\ref{inst:22}},  
   D. Bayliss\inst{\ref{inst:9},\ref{inst:10}},        
   A.~Y. Burdanov\inst{\ref{inst:24}},     
   E. Ducrot\inst{\ref{inst:38},\ref{inst:39}},     
   M. Ghachoui\inst{\ref{inst:21},\ref{inst:37}},     
   M. Gillon\inst{\ref{inst:21}},       
   Y. G\'omez Maqueo Chew\inst{\ref{inst:29}},   
   E. Jehin\inst{\ref{inst:34}},   
   P.~P. Pedersen\inst{\ref{inst:31},\ref{inst:32}},   
   F.~J. Pozuelos\inst{\ref{inst:26}},  
   P.~J. Wheatley\inst{\ref{inst:10},\ref{inst:9}},    
   S. Z\'uniga-Fern\'andez\inst{\ref{inst:21}},       
   Y. Carteret\inst{\ref{inst:1}},   
   H.~M. Cegla\inst{\ref{inst:9},\ref{inst:10}}, 
   A.~C.~M. Correia\inst{\ref{inst:40},\ref{inst:41}}, 
   Y.~T. Davis\inst{\ref{inst:20}},    
   L. Doyle\inst{\ref{inst:10},\ref{inst:9}}, 
   D. Ehrenreich\inst{\ref{inst:1}},        
   N.~C. Hara\inst{\ref{inst:17}}, 
   B. Lavie\inst{\ref{inst:1}},  
   J. Lillo-Box\inst{\ref{inst:11}},  
   C. Lovis\inst{\ref{inst:1}}, 	   
   A.~C. Petit\inst{\ref{inst:3}},
   N.~C. Santos\inst{\ref{inst:4},\ref{inst:13}},   
   M.~G. Scott\inst{\ref{inst:20}},   
   J. Venturini\inst{\ref{inst:1}}, 
   E.-M. Ahrer\inst{\ref{inst:7}}, 
   S. Aigrain\inst{\ref{inst:33}}, 
   S.~C.~C. Barros\inst{\ref{inst:4},\ref{inst:13}},
   E. Gillen\inst{\ref{inst:6}},
   X. Luo\inst{\ref{inst:35},\ref{inst:36}},
   C. Mordasini\inst{\ref{inst:15}},
   K. Al Moulla\thanks{SNSF Postdoctoral Fellow}\inst{\ref{inst:4}},
   F. Pepe\inst{\ref{inst:1}},   
   A.~G.~M. Pietrow\inst{\ref{inst:8}}
   }

   \institute{
   Observatoire Astronomique de l'Universit\'e de Gen\`eve, Chemin Pegasi 51b, CH-1290 Versoix, Switzerland\label{inst:1},   
   \and European Southern Observatory, Alonso de C{\'o}rdova 3107, Vitacura, Region Metropolitana, Chile\label{inst:2}
   \and Universit\'e C\^ote d'Azur, Observatoire de la C\^ote d'Azur, CNRS, Laboratoire Lagrange, France\label{inst:3}
   \and Instituto de Astrof\'isica e Ci\^encias do Espa\c{c}o, Universidade do Porto, CAUP, Rua das Estrelas, 4150-762 Porto, Portugal\label{inst:4}
   \and Astronomy Unit, School of Physics and Astronomy, Queen Mary University of London, London E1 4NS, UK\label{inst:6}
   \and Max-Planck-Institut f\"{u}r Astronomie, K\"{o}nigstuhl 17, 69117 Heidelberg, Germany\label{inst:7}
   \and Leibniz-Institut f\"{u}r Astrophysik Potsdam (AIP), An der Sternwarte 16, 14482 Potsdam, Germany\label{inst:8}
   \and Department of Physics, University of Warwick, Gibbet Hill Road, Coventry CV4 7AL, UK\label{inst:9}
   \and Centre for Exoplanets and Habitability, University of Warwick, Gibbet Hill Road, Coventry CV4 7AL, UK\label{inst:10}
   \and Centro de Astrobiolog\'ia, CSIC-INTA, Camino Bajo del Castillo s/n, 28692 Villanueva de la Ca{\~n}ada, Madrid, Spain\label{inst:11}
   \and Potsdam University, Institute of Physics and Astronomy, Karl-Liebknecht-Str. 24/25, 14476 Potsdam, Germany\label{inst:12}
   \and Departamento de F\'isica e Astronomia, Faculdade de Ci\^encias, Universidade do Porto, Rua do Campo Alegre, 4169-007 Porto, Portugal\label{inst:13}
   \and International Center for Advanced Studies (ICAS) and ICIFI (CONICET), ECyT-UNSAM, Campus Miguelete, 25 de Mayo y Francia, (1650) Buenos Aires, Argentina\label{inst:14}
   \and Space Research and Planetary Sciences, Physics Institute, University of Bern, Gesellschaftsstrasse 6, 3012 Bern, Switzerland\label{inst:15}
   \and Center for Space and Habitability, University of Bern, Gesellschaftsstrasse 6, 3012 Bern, Switzerland\label{inst:16}
   \and Aix Marseille Universit{\'e}, CNRS, CNES, LAM, Marseille, France\label{inst:17}
   \and Center for Astrophysics | Harvard \& Smithsonian, 60 Garden St, Cambridge, MA 02138, USA\label{inst:18}
   \and Univ. Grenoble Alpes, CNRS, IPAG, F-38000 Grenoble, France\label{inst:19}
   \and School of Physics \& Astronomy, University of Birmingham, Edgbaston, Birmingham B15 2TT, UK\label{inst:20}
   \and Astrobiology Research Unit, Universit\'e de Li\`ege, All\'ee du 6 Aout 19C, B-4000 Li\`ege, Belgium\label{inst:21}
   \and Instituto de Astrof\'isica de Canarias (IAC), 38205 La Laguna, Tenerife, Spain\label{inst:22}
   \and Departamento de Astrof\'isica, Universidad de La Laguna (ULL), 38206, La Laguna, Tenerife, Spain\label{inst:23}
   \and Department of Earth, Atmospheric and Planetary Sciences, Massachusetts Institute of Technology, Cambridge, MA 02139, USA\label{inst:24}
   \and Instituto Tecnol\''ogico de Buenos Aires (ITBA), Iguaz\'u 341, Buenos Aires, CABA C1437, Argentina\label{inst:25}
   \and Instituto de Astrof\'isica de Andaluc\'ia (IAA-CSIC), Glorieta de la Astronom\'ia s/n, 18008 Granada, Spain\label{inst:26}
   \and D\'epartement de Physique, Institut Trottier de Recherche sur les Exoplan\`etes, Universit\'e de Montr\'eal, Montr\'eal, Qu\'ebec, H3T 1J4, Canada\label{inst:27}
    \and Department of Physics and Astronomy, McMaster University, 1280 Main St W, Hamilton, ON L8S 4L8, Canada\label{inst:28}
    \and Instituto de Astronom\'ia, Universidad Nacional Aut\'onoma de M\'exico, Ciudad Universitaria, 04510 Ciudad de M\'exico, M\'exico\label{inst:29}
    \and Mullard Space Science Laboratory, University College London, Holmbury St Mary, Dorking, RH5 6NT, UK\label{inst:30}
    \and Cavendish Laboratory, JJ Thomson Avenue, Cambridge CB3 0HE, UK \label{inst:31}
    \and Institute for Particle Physics and Astrophysics , ETH Z\"urich, Wolfgang-Pauli-Strasse 2, 8093 Z\"urich, Switzerland \label{inst:32}
    \and Oxford Astrophysics, Denys Wilkinson Building, Department of Physics, University of Oxford, OX1 3RH, UK\label{inst:33}
    \and STAR Institute, University of Li\`ege, All\'ee du 6 ao\^ut, 19, 4000 Li\`ege (Sart-Tilman), Belgium\label{inst:34}
    \and Max Planck Institute for Astronomy (MPIA), K\"{o}nigstuhl 17, 69117 Heidelberg, Germany\label{inst:35}
    \and School of Physics and Astronomy, China West Normal University, Nanchong 637009, People's Republic of China\label{inst:36}
    \and Oukaimeden Observatory, High Energy Physics and Astrophysics Laboratory, Faculty of sciences Semlalia, Cadi Ayyad University, Marrakech, Morocco\label{inst:37}
    \and LESIA, Observatoire de Paris, CNRS, Universit{\'e} Paris Diderot, Université Pierre et Marie Curie, Meudon, France\label{inst:38}
    \and Universit{\'e} Paris-Saclay, Universit\'e Paris Cit\'e, CEA, CNRS, AIM, Gif-sur-Yvette, France\label{inst:39}
    \and CFisUC, Departamento de F{\i}sica, Universidade de Coimbra, 3004-516 Coimbra, Portugal\label{inst:40}
    \and IMCCE, UMR 8028 CNRS, Observatoire de Paris, PSL Universit\'e, 77 Avenue Denfert-Rochereau, 75014 Paris, France\label{inst:41}
    \and  Instituto de Astronom\'ia, Universidad Cat\'olica del Norte, Angamos 0610, 1270709, Antofagasta, Chile\label{inst:42}
   }

    \date{Received dd:mm:yyyy / Accepted dd:mm:yyyy}

\authorrunning{V. Bourrier et al.}

\offprints{V.B. (\email{vincent.bourrier@unige.ch})}


  \abstract{The distribution of close-in exoplanets is shaped by a complex interplay between atmospheric and dynamical processes. The Desert, Ridge, and Savanna (respectively a lack, overoccurence, and mild deficit of Neptunes with increasing periods) illustrate the sensitivity of these worlds to such processes, making them ideal targets to disentangle their roles. Determining how many Neptunes are brought close-in by early disk-driven migration (DDM, expected to maintain primordial spin-orbit alignment) or late high-eccentricity tidal migration (HEM, expect to generate large misalignments) is essential to understand how much atmosphere they lost. In this paper we propose a unified view of the exo-Neptunian landscape to guide its exploration, speculating that the Ridge is a hotspot for evolutionary processes. Low-density Neptunes would mainly undergo DDM, getting fully eroded at shorter periods than the Ridge, in contrast to denser Neptunes that would be brought to the Ridge and Desert by HEM. We embark on this exploration via the ATREIDES (Ancestry, Traits, and Relations of Exoplanets Inhabiting the Desert Edges and Savanna) collaboration, which relies on spectroscopic and photometric observations of $\sim$60 close-in Neptunes, their reduction with robust pipelines, and their interpretation with internal structure, atmospheric, and evolutionary models. We carry out a systematic Rossiter-McLaughlin census with the VLT/ESPRESSO to measure the distribution of 3D spin-orbit angles, correlate its shape with the system properties (orbit, density, evaporation), and thus relate the fraction of aligned / misaligned Neptunian systems to DDM, HEM, and atmospheric erosion. The first ATREIDES target, TOI-421~c, lies in the Savanna with a neighbouring sub-Neptune TOI-421~b. We measure for the first time their 3D spin-orbit angles ($\psi_\mathrm{b}$ = 57$\stackrel{+11}{_{-15}}$ $^{\circ}$; $\psi_\mathrm{c}$ = 44.9$\stackrel{+4.4}{_{-4.1}}$ $^{\circ}$). Together with the eccentricity and possibly large mutual inclination of their orbits, this hints at a chaotic dynamical origin that could result from DDM followed by HEM. Our program will provide the community with a wealth of constraints for formation and evolution models, and we welcome collaborations that will contribute pushing forward our understanding of the exo-Neptunian landscape.}

   \titlerunning{The ATREIDES collaboration I}
   \maketitle
%

\section{Introduction} 
\label{sec:introduction}

\begin{figure}
\includegraphics[trim=2cm 0.5cm 5cm 1cm,clip=true,width=\columnwidth]{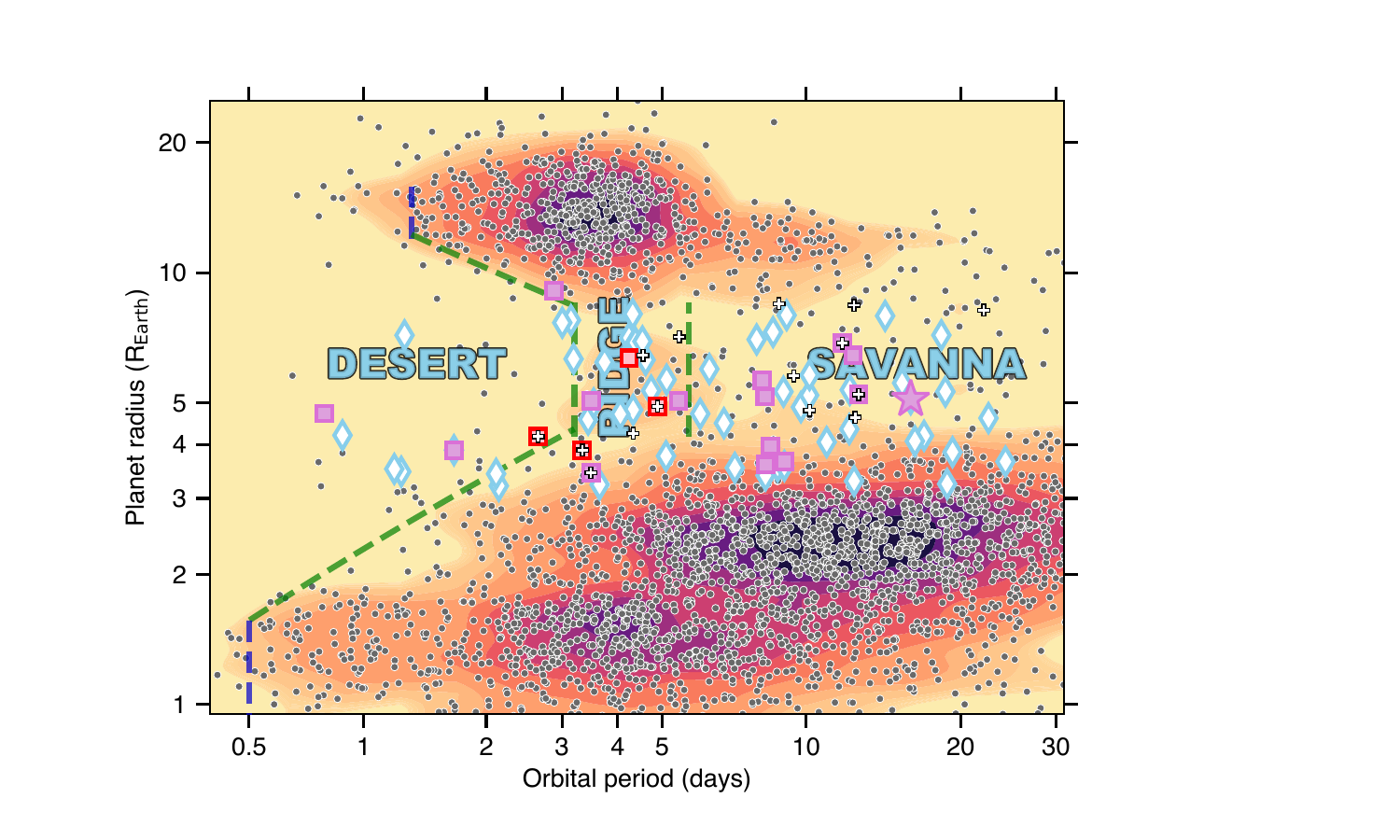}
\centering
\caption[]{Radius of close-in exoplanets (known to better than 30\% precision) as a function of orbital period, based on the NASA Exoplanet Archive Composite database\protect\footnotemark[2]. Background color scales with observed occurrence, increasing from yellow to black. Dashed lines show the Desert and Ridge boundaries (\citealt{CastroGonzalez2024a}). Diamonds indicate ATREIDES targets, with a star for TOI-421~c. Square symbols show Neptunes for which escaping H/He was detected (red; GJ 436~b, GJ 3470 b, HAT-P-26~b, HAT-P-11 b from left to right) or not (magenta). Crosses show Neptunes for which the 3D spin-orbit angle has been measured
(\textit{Tepcat}\protect\footnotemark[3]) to better than 30$^{\circ}$. Data were retrieved in May 2025.} 
\label{fig:Per_Rad_PAPER}
\end{figure}

The discovery of 51\,Peg b \citep{Mayor1995} and the realization that giant planets exist on short orbits changed the paradigm derived from our interpretation of the Solar System \citep[e.g.][]{Lin1996,Rasio1996}. 30 years later, more than 60\% of known exoplanets\footnote{From \url{https://exoplanet.eu/home/} and \url{https://exoplanetarchive.ipac.caltech.edu/}} have been discovered on orbits shorter than 30\,days, confirming the prevalence of these so-called ``close-in'' planets and their importance for planetology (\citealt{Lovis2009,Howard2012,Fressin2013}). The landscape of close-in planets is far from being homogeneous. A lack of Neptune-size planets has been mapped over the last fifteen years (Fig.~\ref{fig:Per_Rad_PAPER}), emerging as a ``Desert'' in which exist a rare few hot Neptunes (R$_{\rm p}\sim 2 - 10$ $R_{\oplus}$; P $\approxinf$ 3\,d; e.g., \citealt[]{Lecav2007,Davis2009,BenitezLlambay2011,Szabo2011,Youdin2011,Beauge2013,Lundkvist2016,Mazeh2016}) and a mildly populated ``Savanna'' of warm Neptunes at longer periods (\citealt{Bourrier2023}; R$_{\rm p}\sim 4 - 9$ $R_\oplus$; P $\approxsup$ 3\,d). The Desert and Savanna are not observational biases, since we have the capability to detect planets in this size and mass range up to periods beyond 30\,d. Recently, \citealt{CastroGonzalez2024a} computed the underlying distribution of close-in planets to derive bias-corrected boundaries of the Desert, and found that it is separated from the Savanna by a ``Ridge'', an overdensity of Neptunes within 3.2--5.7\,days (Fig.~\ref{fig:Per_Rad_PAPER}). This complex landscape bears the imprint of the various processes that shaped the exoplanet population, making close-in Neptunes targets of choice to study planetary formation and evolution.

\footnotetext[2]{\url{https://exoplanetarchive.ipac.caltech.edu/cgi-bin/TblView/nph-tblView?app=ExoTbls&config=PSCompPars}}
\footnotetext[3]{\url{https://www.astro.keele.ac.uk/jkt/tepcat/obliquity.html}}

Runaway gas accretion predicts a drought of Neptune-mass planets when compared to Jupiter-mass planets, as cores massive enough to trigger accretion within the protoplanetary disk swiftly grow to Jupiter-mass if there is enough available material \cite[e.g.,][]{Mordasini2015,Batygin2016}. Yet, this process is not efficient interior of the ice line (the orbital distance beyond which water can condense), where the disk contains less solid material and cores typically do not reach a sufficient mass. This means that Jupiter- and Neptune-size planets migrate close to their star after forming beyond the ice line \cite[e.g.,][]{Rafikov2006,Dawson2018}, and that runaway accretion may not be directly responsible for the observed deficit of close-in Neptunes. Indeed, whether the orbital distribution of close-in Neptunes actually reproduces their primordial distribution beyond the ice line is an open question, related to the relative roles of the different migration processes and the final orbits to which they bring Neptune-size worlds. 

Smooth, disk-driven migration (DDM, e.g. \citealt{Goldreich1979,Lin1996,Baruteau2016}) occurs early-on and is expected to preserve the spin-orbit angle acquired upon formation \cite[e.g.,][]{Palle2020b,Zhou2020,Mann2020}. It either traces the primordial alignment between star and disk inherited from the collapsing molecular cloud, oderate misalignment (e.g.\citealt{Albrecht2022,Takaishi2020}) inherited from the star (chaotic formation, e.g. \citealt{Bate2010,Thies2011,Fielding2015,Bate2018}; internal gravity waves, e.g. \citealt{Rogers2012,Rogers2013}; magnetic torques, e.g. \citealt{Lai2011}; gravitational torques from companions, e.g. \citealt{Tremaine1991,Batygin2011,Storch2014}) or from the disk \cite[e.g.,][]{Foucart2011,Batygin2012,Lai2014,Zanazzi2018, Romanova2021}. In contrast, high-eccentricity tidal migration (HEM, such as planet--planet scattering, e.g.  \citealt{Ford2008,Chatterjee2008,Nagasawa2008,Nagasawa2011,Beauge2012,Gratia2017}; Kozai--Lidov migration, e.g.  \citealt{Wu2003,Correia2011,Naoz2011,Teyssandier2013}; secular chaos, e.g. \citealt{Wu2011}) can occur at any time, from just after the disk dissipation up to several billion years, and is expected to induce highly misaligned orbits with no memory of the primordial orientation \citep[e.g.,][]{Naoz2012,Albrecht2012,Nelson2017}. The present-day angle between the spin of a star and the normal to its planets' orbital plane is thus an essential observational tracer to distinguish between DDM and HEM. The sky-projected spin-orbit angle can be measured through the distortion of the stellar lines induced by transiting planets (the Rossiter-McLaughlin, or RM, signal, \citealt{Rossiter1924,McLaughlin1924} and review by \citealt{Triaud2018}), which traces the orientation of their orbit and has been analyzed with increasingly refined techniques (RV anomaly, \citealt{queloz2010}; Reloaded RM, \citealt{Cegla2016}; RM Revolutions, \citealt{Bourrier2021}). 

The evolution of close-in Neptunes likely depends on the specific interplay between each migration pathway and evaporation \citep{Bourrier_2018_Nat,Owen2018,Attia2021,Vissapragada2022,CastroGonzalez2024a}. Hydrodynamical escape induced by the stellar X-ray and extreme-UV irradiation is thought to sculpt the Desert \citep[e.g.,][]{Lecav2004,Owen2012,Owen2019,McDonald2019} by removing volatile species from Neptune-size planets and eroding them into mini-Neptunes and bare rocky cores \citep[e.g.,][]{Ehrenreich_desert2011,Lopez2013,Pezzotti2021}. However, it is yet to be determined whether the transition from quiescent to hydrodynamical escape \citep[e.g.,][]{Koskinen2007} occurs within the Desert, Ridge, or further out into the Savanna, and at what stage of their life Neptunes evaporate. Indeed, it is usually assumed that the bulk of atmospheric erosion occurs early-on, during formation or just after DDM \cite[e.g.,][]{Jin2014}. However, Neptunes undergoing late HEM, which has also been proposed as one of the processes shaping the Desert \citep[e.g.,][]{Matsakos2016, Mazeh2016,Owen2018}, would not be subjected to the strong saturated XUV irradiation phase from their young host star (\citealt{Bourrier_2018_Nat,Attia2021}). 

Determining the fraction of Neptunes brought close-in by early DDM or late HEM is thus critical to understand their overall evolution. This requires building a complete distribution of spin-orbit angles across the exo-Neptunian landscape, which is the primary goal of the Ancestry, Traits, and Relations of Exoplanets Inhabiting the Desert Edges and Savanna (ATREIDES) collaboration. After proposing a comprehensive view of the exo-Neptunian landscape and introducing the collaboration (Sect.~\ref{sec:landscape}), we describe our observational (Sect.~\ref{sec:obs_strategy}), data reduction (Sect.~\ref{sec:red_strategy}), and analysis (Sect.~\ref{sec:ana_strategy}) strategies. We present the first representative result of ATREIDES (Sect.~\ref{sec:TOI421}) and our perspectives for the publication series (Sect.~\ref{sec:discuss}).


\section{Exploring the exo-Neptunian landscape} 
\label{sec:landscape}

\subsection{Toward a unified view} 
\label{sec:theory}

We propose a comprehensive view of the exo-Neptunian landscape, attempting to integrate the roles of photo-evaporation, DDM, and HEM. It is inspired by previous studies (e.g. \citealt{Bourrier_2018_Nat,Owen2018,Kubyshkina2022}) and our analysis of critical properties measured for the close-in Neptune population (Fig.~\ref{fig:prop_nept}). In particular, based on the density dichotomy identified by \citet{CastroGonzalez2024b}, we divide the Neptunian population between low-density (``fluffy'') and high-density (``dense'') classes separated by $\rho\sim$1\,g\,cm$^{-3}$. We outline here the main conclusions derived from our reasoning, fully developed in Appendix~\ref{apn:nept_orig}. We caution that they remain first-order hypotheses, intended to guide and be validated by more detailed theoretical and modelling studies:
\begin{itemize}[leftmargin=*]
    \vspace{-0.5\baselineskip}
    \item Fluffy and dense Neptunes do not appear related through evaporation. This implies that they formed as two primordial populations distinct in density, which determined their dynamical and atmospheric evolution.
    \item DDM may distribute Neptunes over various orbital periods at an early age, acting primarily on fluffy Neptunes. It would preserve their primordial near-circular, aligned orbits. 
    \item HEM may act primarily on dense Neptunes, bringing them preferentially to the Ridge or to the Desert depending on their core and atmospheric mass. They would end up on mildly-eccentric and misaligned orbits, which could then be further excited by additional dynamical processes within the Ridge depending on system-specific properties (e.g., the primordial architecture and mass ratios between the star, planet, and companion). In particular, we speculate that some systems are brought to a stable ``polar'' configuration, rather than being tidally-realigned. 
    \item The lower envelope of the density distribution can be refined into a ``brink'' within the Ridge, increasing with decreasing orbital periods. We propose it marks the threshold for full atmospheric erosion of fluffy Neptunian planets migrating early-on. Only dense Neptunes, or late Neptunian migrators, would survive within the Desert.
    \item The similarity in occurrences, eccentricities, and misalignments between the Neptunian and Jupiter landscapes suggests they are shaped by similar processes, albeit with different efficiencies.
\end{itemize}

It is noteworthy that the four known evaporating Neptunes (GJ 436 b, \citealt{Butler2004}; HAT-P-11 b, \citealt{Bakos2010}; GJ 3470~b, \citealt{Bonfils2012}; HAT-P-26 b, \citealt{Hartman2011a}) are located within the Ridge. GJ 436 b, HAT-P-11 b, and GJ 3470~b are on eccentric and polar orbits (\citealt{Bourrier_2018_Nat,Stefansson2021,Bourrier2022,Bourrier2023}), thus they likely underwent HEM followed by the dynamical ``polarization'' process pumping misaligned orbits to a spin-orbit angle of 90$^{\circ}$. The orbital architecture of HAT-P-26 b is not known precisely, but since it is also on an eccentric orbit within the Ridge we predict that it underwent HEM and ended up on a mildly misaligned or polar orbit, depending on its sensitivity to the polarization process. GJ 436 b and HAT-P-11 b are dense Neptunes with moderate mass loss rates (e.g. \citealt{Ehrenreich2015,Bourrier2016,Allart2018,BenJaffel2022}), and we do not expect their evolution to have been heavily impacted by erosion (Attia et al., in prep). GJ 3470 b and HAT-P-26 b are fluffy Neptunes with mass loss rates strong enough that they would have eroded half of their primordial mass if they arrived early-on (\citealt{Bourrier2018_GJ3470b,Vissapragada2022}). They may thus owe their survival to a late migration, and their present location near the density brink further suggests that this feature is also partly shaped by HEM. 

If the proposed hypotheses hold, we expect the following:
\begin{itemize}[leftmargin=*]
    \vspace{-0.5\baselineskip}
    \item Ridge Neptunes, especially fluffy ones along the density brink, would be prime targets to detect evaporation.
    \item At very short orbital distances the orbits of Neptunes should get reshaped by tides and their atmosphere be partially or fully eroded. The detection within the Desert of a fluffy Neptune, or a dense Neptune on an eccentric and misaligned orbit, would then be a smoking gun for recent migration.
    \item The distribution of spin-orbit angles should resolve into aligned or mildly-misaligned within the Savanna, and into aligned or polar within the Ridge. Polarized Neptunes, being stable (\citealt{Louden2024}), would keep the trace of HEM processes, unlike tidally re-aligned systems which would be indistinct from primordial alignments.
\end{itemize}

Reality is certainly more complex than the proposed scenarios. We remain limited by the lack of critical properties to explore in detail Neptunian evolution, in particular mass loss and orbital architecture measurements, motivating large-scale surveys of these properties.


\subsection{The ATREIDES collaboration} 
\label{sec:collab}

The ATREIDES collaboration is built upon a large program (PI: V. Bourrier; Prog. ID: 112.25BG.00N) obtained on the VLT/ESPRESSO (\citealt{Pepe2021}) to probe Neptunian orbital architectures. ATREIDES proceeds from the DREAM pathfinder program (\citealt{Bourrier2023}), which targeted a sample of 14 close-in planets from mini-Neptune to Jupiter-size along the rim of the Desert. To mitigate selection biases in ATREIDES, we defined a volume-limited sample in period-radius space encompassing all close-in (P $<$ 30\,d) Neptunes (3.2\,R$_{\oplus}<$ R$_{\rm p}<$ 8.5\,R$_{\oplus}$). A period of 30\,d was chosen as a typical definition for close-in planets, covering the Desert, Ridge, and Savanna, and because there are few transiting Neptunes known at longer periods and observable within a single night. We further selected planets with unknown or poorly known spin-orbit angles, and a RM Revolutions (RMR, \citealt{Bourrier2021}) signal detectable with ESPRESSO in a single night. This is based on a first-order criterion for the detectability of the planet-occulted line, so that ATREIDES results will be used to calibrate RM detection limits and the parameter space where finer stellar effects can be detected (e.g., \citealt{RoguetKern2022}). The current sample is shown in Fig.~\ref{fig:Per_Rad_PAPER}. Targets may be replaced as better ones are identified, and the sample may grow as targets from other programs are included.

The main objectives of ATREIDES are:
\begin{itemize}[leftmargin=*]
    \vspace{-0.5\baselineskip}
    \item to increase substantially spin-orbit angle measurements for close-in Neptunes, in 3D whenever possible.
    \item to determine the frequency of aligned/misaligned systems, and more generally the distribution of spin-orbit angles and their dependence with planet and star properties
    \item to infer the roles of DDM and HEM in shaping the Neptunian population, and their interplay with evaporation.
\end{itemize}

Secondary objectives include characterizing stellar and planetary atmospheres, and probing the irradiation threshold for the onset of hydrodynamical escape and structural changes in the planetary atmosphere. An important by-product of ATREIDES is the revision of stellar properties and planetary ephemeris.

Our observational strategy (Sect.~\ref{sec:obs_strategy}) is driven by the detectability of the RM signals and visibility of our targets. To complement the ESPRESSO observations, we gained access to photometric facilities to revise the planet ephemeris and better characterize their host stars. As the reproducibility and robustness of scientific results is essential to derive meaningful conclusions, ATREIDES and archival ESPRESSO data relevant to our goals will be analyzed homogeneously by applying standard reduction (Sect.~\ref{sec:red_strategy}) and interpretation tools (Sect.~\ref{sec:ana_strategy}).


\section{Observational strategy}
\label{sec:obs_strategy}

In this section, we provide detailed information about our observational strategy, so that it can benefit future RM observing programs.

\subsection{Transit spectroscopy}

\subsubsection{ESPRESSO observations}

ATREIDES transits are observed with the ESPRESSO spectrograph (\citealt{Pepe2021}), installed on the Very Large Telescope (VLT) at ESO's Paranal site. ESPRESSO disperses light on 85 spectral orders from 380 to 788\,nm with a resolving power of 140~000. The small transit depth of Neptune-size planets limited their RM measurements to only a few objects (e.g., \citealt{Bourrier_2018_Nat,Wang2018_Kepler9,Zhou2018,Gaidos2020,Stefansson2022,Wirth2021}) until the VLT large collecting area, ESPRESSO high spectral resolution, and new RM techniques (\citealt{Cegla2016,Bourrier2021}) expanded the field toward smaller planets, fainter hosts, and stars with slowler rotation (e.g., \citealt{Bourrier2023}). Combined with more discoveries of Neptune-sized planets transiting bright stars from TESS \citep{Ricker2015}, this allowed us to define a sample of about 60 close-in Neptunes amenable to RM measurements.

All observations are carried out using ESPRESSO mode HR21 to reduce read-out noise, and with fiber B on the sky to monitor possible contamination. Analyzing datasets obtained with the same spectrograph and observational settings strengthens the homogeneity and comparability of the derived properties. We chose to observe each planet's transit once, as RM signals are less prone to variability than planetary atmospheric signatures. Planets yielding RM signals at lower S/N than expected can be re-observed in the future, combining new and archival data in a common analysis. If possible, each transit is observed with baseline both before ingress and after egress to build a reference spectrum for the unocculted star, and to characterize short-term variations due to the star, the Earth's atmosphere, and the instrument, which can then be corrected for over the entire dataset. Transit durations of all ATREIDES targets are short enough that they can each be observed within a single night, but it is challenging to observe the farthest-orbiting ones with long baselines. Observations of Neptunes on periods longer than 30 days, typically with longer and less frequent transits, will be more difficult to observe in future programs and may require observations acquired on different nights.

Using a custom-made transit planner\footnote{PTO, publicly available at \url{https://gitlab.unige.ch/spice_dune/pto}}, we identified all transit windows observable in good conditions for each planet, that is with transit and baseline observable within a single night at airmass lower than 2.2, if possible outside of twilight, and with no or limited moon contamination (moon at more than 45$^{\circ}$ from the target). ATREIDES is the first large program at the VLT composed entirely of transit observations, which raised new scheduling challenges compared to traditional radial velocity (RV) programs, for which observations are not as time-critical. Planets with more than three observable windows in a given semester could be observed by ESO in service mode, while planets with fewer opportunities are observed in designated visitor mode by ATREIDES collaborators.

Transit observations are made of consecutive exposures obtained with high temporal cadence. For each planet, we determined the smallest exposure time to obtain a S/N per pixel of the stellar spectrum $>$30 at 550\,nm for a seeing of 1'' (so that spectra are properly reduced by the ESPRESSO DRS), to ensure a favorable ratio between science and overhead time during an exposure, and to smooth out high-frequency stellar variability.

We checked that these exposure times yielded at least 6 in-transit exposures, to ensure a spatial sampling of the transit chord fine enough to constrain the orientation of the planetary orbit. Final exposure times range between 180 and 500\,s for most ATREIDES systems, extending up to 1500\,s in a few cases while still allowing successful RM signal extractions.

\subsubsection{Ephemeris revision}
\label{sec:ana_ephem}

RM analyses require precise transit ephemerides to avoid cross--contamination between the in-transit signal and stellar baseline, and to position the RM model at the correct orbital phases. For high-quality RM signals, even mid-transit time ($T_{0}$) uncertainties of the order of a minute can bias spin-orbit angle measurements significantly (e.g., \citealt{CasasayasBarris2021}). Unfortunately, the precision on $T_{0}$ degrades quickly for planets with close-in orbits and imprecise periods. For each ATREIDES planet, we selected the published ephemeris that provided the most precise $T_{0}$ propagated at the epoch of the ESPRESSO observations. However, the ephemeris of many planets detected several years ago were not updated, resulting in uncertainties up to several hours. We thus conducted an extensive campaign with ground-based photometric facilities (Sect.~\ref{sec:obs_photom}), combined with a re-analysis of archival data and new TESS data (when available and precise enough), to get below 5~min precision on the propagated $T_{0}$. This threshold was chosen as a compromise between biases on the RM interpretation (considering ATREIDES signals are measured at lower S/N than in \citealt{CasasayasBarris2021}) and the number of transit observations required to get to such precisions. 

Due to the numerous targets, heterogeneous datasets, and time-sensitive need to determine optimal transit windows, we developed a semi-automated transit analysis pipeline. This approach allowed us to quickly revise ephemerides when new photometry was obtained and to assess the need for further follow-up. We describe the pipeline, its standard operation, and current results in Sect.~\ref{sec:photom_strat}. Final ephemerides will be derived in the relevant ATREIDES publications based on more thorough analyses.

\subsection{Photometry}
\label{sec:obs_photom}

The ATREIDES program relies on photometry for three main objectives: to revise the planet ephemerides with recent transit observations, to catch spot crossings simultaneous to the ESPRESSO transits, and to monitor the activity and rotation of the star with long-term (days-to-years) observations. The first objective is essential for scheduling purposes and to properly interpret the ESPRESSO observations. The second can bring useful information on the stellar photosphere, using the planet as a probe, and can help better constrain the system architecture. The third allows us to disentangle systematics due to the star from planetary signatures, and to break the degeneracy between the stellar rotational velocity and inclination (thus yielding the 3D spin-orbit angle from the analysis of the RM signal, Sect.~\ref{sec:RMR}). To achieve these goals ATREIDES includes collaborators with access to the CHEOPS, Euler, MuSCAT2, NGTS, SPECULOOS, STELLA, and TRAPPIST facilities. We further exploited photometric data obtained outside of ATREIDES, and available through public access or collaborations.

\subsubsection{Dedicated photometry}
\label{sec:obs_dedicated_photometry}

The CHaracterising ExOplanets Satellite (CHEOPS) is a 30-cm  photometric space telescope dedicated to the study of known exoplanets via high precision time-series photometry \citep{Benz2021}. CHEOPS was launched on 18 December 2019 and orbits Earth in a 700\,km altitude sun-synchronous polar orbit with a period of 100 minutes. It is a pointed mission, observing one source at the time through a broad optical bandpass. The instrument is defocused to optimise precision for bright stars, and CHEOPS has repeatedly shown to reach photometric precisions better than 10 ppm over timescales of several hours for bright stars \citep[e.g.][]{Lendl2020,Delrez2021}. For details on the in-flight performance of CHEOPS, please refer to \citet{Fortier2024}. The standard CHEOPS data reduction pipeline  \citep{Hoyer2020} uses aperture photometry, providing fluxes for a range of aperture sizes together with instrument state variables that can be used as cotrending basis vector during the data analysis, which is often done using the open-source tool \emph{PyCHEOPS} \citep{Maxted2023}. As an alternative to the standard data reduction pipeline, an independent open-source PSF-based photometric extraction software, \emph{PIPE}, is also available \citep{Brandeker2024}. We collaborate with the CHEOPS consortium for specific targets benefiting from high-precision photometry. 

EulerCam is a 4k$\times$4k CCD imager installed at the 1.2\,m Euler telescope of La Silla observatory. It has a field of view (FoV) of 14.7$\times$14.7\,arcmin$^{2}$ and a pixel scale of 0.215 arcsec per pixel. The instrument is equipped with the full set of Geneva filters \citep{Rufener1988}, an \emph{I-Cousins} filter, \emph{r'-} and \emph{z'-}Gunn filters and a broad (520-880\,nm) \emph{NGTS} filter. We exploit the large collecting area of Euler mainly to perform simultaneous transit observations, usually using a \emph{V$_G$} or \emph{r'} filter to remain sensitive to stellar surface inhomogeneities while reaching a high photometric precision. The instrument and the associated data reduction procedures are described in detail in \citet{Lendl2012}. In short, relative aperture photometry is performed on the target using an iteratively-chosen set of bright nearby references, with the extraction aperture and reference star selection optimised to achieve the minimal light curve RMS. 

MuSCAT2 \citep{Narita2019} is a multi-band imager mounted on the 1.5\,m Telescopio Carlos S\'{a}nchez (TCS) at the Teide Observatory, Spain. The instrument is equipped with four CCDs and can obtain simultaneous images in the $g'$, $r'$, $i'$, and $z_s$ bands with a short readout time. Each CCD has 1024$\times$1024 pixels with a FoV of 7.4$\times$7.4\,arcmin$^2$. Exposure times can be set independently for each filter, allowing the optimal peak counts to be set for the target and reference stars in different bands. The raw data have been reduced by the MuSCAT2 pipeline \citep{Parviainen2019}; the pipeline performs dark and flat-field calibrations, aperture photometry, and transit model fitting including the effect of instrumental systematics. MuSCAT2 produces high-precision multicolor time-series to refine the ephemerides of ATREIDES targets. 

The NGTS (Next Generation Transit Survey; \citealt{Wheatley2018}) facility comprises twelve 20\,cm telescopes, situated at the ESO Paranal Observatory in Chile. Each telescope is equipped with f/2.8 astrographs fitted with back-illuminated deep-depletion 2048$\times$2048 pixels CCD and a field-of-view of 8 square degrees, providing a wider range of sufficient reference stars, and observes in a custom filter in the red-optical range of 520--890\,nm. Observations can be acquired in a single telescope survey mode or multiple cameras can concurrently observe to achieve a higher precision. In practice, NGTS is capable of achieving a photometric precision of 400\,ppm in 30\,minutes for stars fainter than $G=12$\,mag (\citealt{Bryant2020}). The combination of multi-camera observations also allows for a reduction of scintillation noise and an improved precision of 100\,ppm in 30\,minutes if all twelve telescopes are used simultaneously on bright targets, as it creates a total effective collecting area equivalent to that of a 0.7\,m telescope and the scintillation noise of a 1.3\,m telescope (\citealt{Bayliss2022}). NGTS contributes to the three photometry objectives, and led the effort of refining the transit ephemerides of ATREIDES planets with high-precision time-series photometric data. The light curves were extracted using the standard NGTS aperture photometry pipeline detailed in \cite{Wheatley2018}.

The SPECULOOS \citep[Search for habitable Planets EClipsing ULtra-cOOl Stars,][]{Delrez2018,Sebastian2021} network is made of six robotic telescopes spread across the globe. The SPECULOOS-Southern Observatory is located at ESO Paranal Observatory in Chile and hosts four robotic 1.0-m Ritchey-Chr{\'e}tien telescopes \cite{Jehin2018Msngr}. The SPECULOOS-North facilities include the SPECULOOS-Northern Observatory located at the Teide Observatory in the Canary Islands in Spain \citep{Burdanov2022}, and the SAINT-Ex \citep[Search And characterIsatioN of Transiting EXoplanets][]{Demory2020} telescope at the National Astronomical Observatory of Mexico in San Pedro M{\'a}rtir. All telescopes are equipped with a 2k$\times$2k deep-depletion Andor CCD camera optimised for the near-infrared up to 1$\mu$m. The pixel scale is 0.35 arcsec which provides a FoV of $12\times12$\,arcmin$^{2}$. The SPECULOOS data are seeing-limited time-series photometry. These facilities produce high-precision time-series photometric data which are used to support the ATREIDES program both in the refinement of ephemerides as well as simultaneous observations with ESPRESSO.  

The STELLA (short for STELLar Activity) observatory is a robotic observing facility located at Iza\~{n}a observatory in Tenerife, Spain,  consisting of two 1.2\,m telescopes named STELLA-I and STELLA-II \citep{Strassmeier2004}. STELLA-I is equipped for photometry with the wide-field imager WiFSIP \citep{Granzer2010, Weber2012}, which covers a FoV of 22'x22' at a scale of 0.32"/pixel. Its detector is a single 4096x4096 back-illuminated thinned CCD with 15 $\mu$m pixels and a peak quantum efficiency of $>90$\%. STELLA-WiFSIP is in principle able to detect shallow transits, but only under excellent observing conditions \citep{Mallonn2022}. Its main use in the ATREIDES program is to characterize stellar variability and measure rotation periods.

The TRAPPIST (TRAnsiting Planets and PlanetesImals Small Telescope) facilities are comprised of two robotic 0.6-m telescopes located at ESO La Silla Observatory in Chile \citep{Gillon2011,Jehin2011} and Ouka{\"i}meden Observatory in Morocco \citep{Barkaoui2019}. They are identical Ritchey-Chr{\'e}tien telescopes with F/8, and are equipped with a German equatorial mount. The camera of TRAPPIST-South is a 2K$\times$2K thermoelectrically cooled FLI ProLine CCD camera with a pixel scale of 0.64 arcsec providing a FoV of $22\times22$\,arcmin$^{2}$, while TRAPPIST-North is equipped with a 2k$\times$2k Andor iKon-L CCD camera with a pixel scale of 0.6 arcsec providing a slightly smaller FoV of $20\times20$\,arcmin$^{2}$. These facilities produce high-precision time-series photometric data which are used to support the ATREIDES program both in the refinement of ephemerides as well as simultaneous observations with ESPRESSO.

\subsubsection{Archival photometry}
\label{sec:obs_complementary_photometry} 

To characterize the long-term photometric activity and further refine the ephemerides of our targets we exploit archival data obtained with various ground-based and space-borne facilities. 

The All-Sky Survey for Supernovae \citep[ASAS-SN;][]{Shappee2014} provides daily full-sky photometry through different stations around the world. Each station comprises four 14\,cm-aperture Nikon telephoto lenses connected to 2048$\times$2048 pixels CCD. The survey pipeline \citep{Kochanek2017} performs aperture photometry at any selected coordinates. ATREIDES targets are typically nearby stars with large proper motion, which can cause flux losses \citep[e.g.][]{Trifonov2021}. We thus use the Sky Patrol\footnote{\url{https://asas-sn.osu.edu/}} to extract photometry in different chunks with durations corresponding to sky-projected distances of 1~arsec, which ensures negligible losses \citep[][]{CastroGonzalez2023}. 

The WASP survey instruments and strategy, and the photometric data analysis pipeline are described in detail in \citet{Pollacco2006}. In brief, each WASP installation consists of a single telescope mount holding 8$\times$200\,mm Canon camera lenses fitted with 2k$\times$2k pixels CCD built by Andor Technology. The Northern and Southern sites are located on La Palma and at Sutherland in South Africa. The survey strategy entails a continuous cycling of pointings along a line of constant declination, which for any given star yields a typical cadence of around 8 minutes. A typical field in the WASP survey was observed for approximately 4-5 months per season. At the end of each season the photometric data for each field are collated and merged, and de-trended using the SYSREM algorithm as described in \citet{CollierCameron2006} in order to reduce instrumental systematics. Given the long and uninterrupted observing baselines of ASAS-SN and WASP, these datasets allow us to estimate magnetic cycle timescales for most ATREIDES targets (differentiating long-term RV signals from outer companions), to independently detect and refine known rotation periods (typically in active stars), and to identify new candidate rotation signals (informing follow-up with more precise facilities, Sect.~\ref{sec:obs_dedicated_photometry}).

TESS (\textit{Transiting Exoplanet Survey Satellite}) is an all-sky discovery mission that focuses on a different region of the sky approximately every month \citep{Ricker2014}. The TESS spacecraft hosts four cameras, each with a FoV of 24$^o$ x 24$^o$ and a focal ratio of f/1.4. Since its launch in 2018, TESS has observed over 80 sectors of data (each consisting of $\sim$28 days of observations), covering almost 95\% of the sky.
This wealth of observations means that the vast majority of bright stars now have at least a month of observations, with many being visited every second year. Such a dataset is highly valuable for the ATREIDES program, as it allows for ephemeris of exoplanets discovered decades ago to be updated ahead of ESPRESSO observations, and for extra information to be gleaned about nearly contemporaneous stellar activity and rotation. 

The K2 mission \citep{Borucki2010} was successor to the \textit{Kepler} mission \citep{Howell2014}, following the failure of two of the Kepler satellite's reaction wheels in 2012 and 2013. The spacecraft hosted a single telescope with a 0.95\,m aperture and a FoV of 115 deg$^2$. Due to its reduced pointing capabilities, the K2 mission focused on the ecliptic plane and detected exoplanets within reach of southern ground-based observing facilities, including many ATREIDES targets. Compared to TESS, the K2 mission benefited from the smaller pixel size and improved precision of the Kepler satellite (e.g. \citealt{Battley2021}), providing precise transit timings and a window into weaker stellar activity. Combining K2 observations with recent transit data provided almost a decade of baseline, greatly improving ephemerides.


\section{Data reduction methodology}
\label{sec:red_strategy}

\subsection{Photometry}
\label{sec:photom_strat}

We generalized and semi-automated the procedures in \citet{CastroGonzalez2022} and \citet{Battley2021} to devise an ephemeris pipeline for the ATREIDES collaboration, combining our dedicated (Sect.~\ref{sec:obs_dedicated_photometry}) and complementary  (Sect.~\ref{sec:obs_complementary_photometry}) photometry to reach the required $T_{0}$ precision (Sect.~\ref{sec:ana_strategy}). The pipeline first flattens the photometry of each facility individually to remove low-frequency trends related to stellar activity or uncorrected instrumental systematics. We typically used simple median or bi-weight filters \citep{Hippke2019} and considered cadences or window lengths about an order of magnitude larger than transit durations to avoid transit-shape deformations. No additional processing was typically applied to photometry already systematics-corrected and normalized by some ground-based facility pipelines. We then combined individual photometry into a single light curve, which we fitted with the \citet{Mandel2002} quadratic limb-darkened transit model in \texttt{batman} \citep{Kreidberg2015}. We included an offset and a white-noise (i.e. jitter) term per facility (or per observing window for discontinuous observations) and considered independent limb-darkening coefficients for each photometric filter (parametrized using the uninformative sampling prescription by \citealt{Kipping2013}). We sampled the posterior probability density function (PDFs) of the transit model parameters through a Markov chain Monte Carlo (MCMC) affine-invariant ensemble sampler \citep{Goodman2010,Foreman2013}. Uninformative priors were set on the parameters, except for $P_{\rm orb}$ and $T_{0}$ on which we set relatively wide Gaussian priors based on the most precise published values and associated 5$\sigma$ uncertainties. 

More dedicated analyses were required for a few systems, such as highly active young stars with strong photometric variations on timescales comparable to the planetary transits and low-S/N transits. In those cases we jointly modeled the transit with a Gaussian Process Regression (GP) using a flexible kernel \citep[e.g.][]{FM2017}, which was shown to preserve the shape of low-S/N transit distorted by unknown red noise structures \citep[e.g.][]{Leleu2021b,Damasso2023}. Furthermore, some planets showed transit timing variations (TTVs) large enough to prevent a precise ephemeris determination. 
In that case we modeled individual transits with a specific $T_{0}$, following the procedures described in \citet{Almenara2022} and \citet{Leleu2021}. 

\begin{figure}
    \centering
    \includegraphics[width=0.47\textwidth]{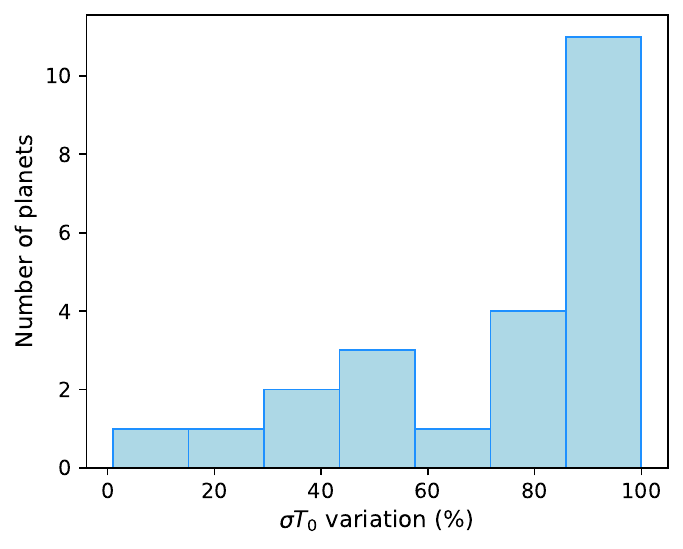}
    \caption{Relative $\sigma T_{0}$ variation of the 23 planets for which we improved the ephemeris to date. The mean variation is 73$\%$.}
    \label{fig:improvement_hist}
\end{figure}

\begin{figure}
    \centering
    \includegraphics[width=0.47\textwidth]{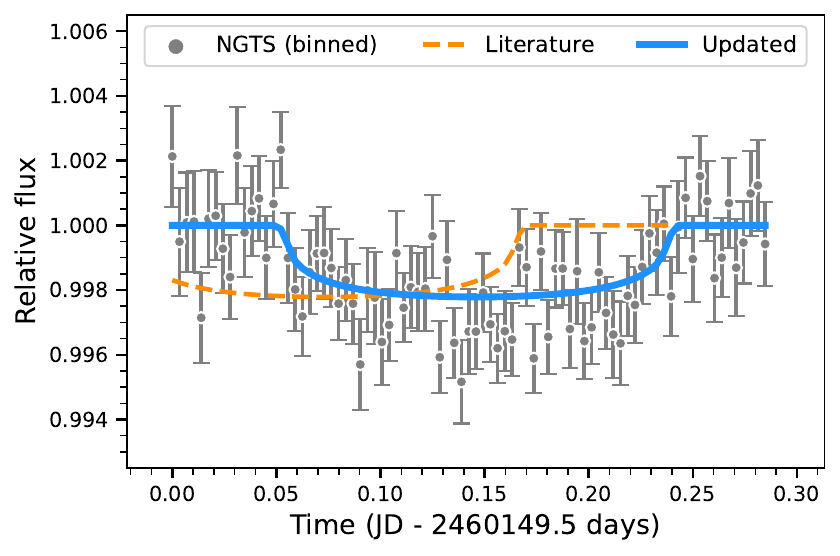}
    \includegraphics[width=0.47\textwidth]{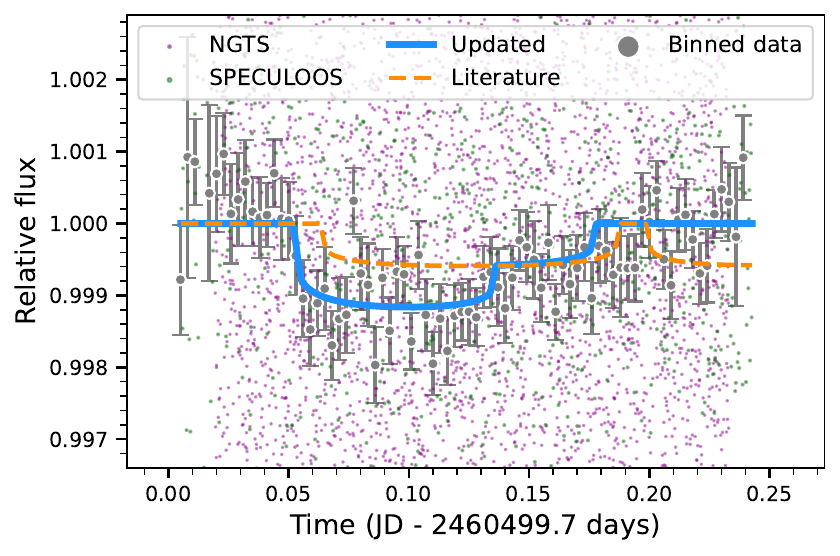}
    \caption{Example transits acquired with NGTS and SPECULOOS (Sect.~\ref{sec:obs_dedicated_photometry}) showing their ephemeris constraining capacity. }
    \label{fig:transits}
\end{figure}

We ran our ephemeris pipeline on 45 planets, improving the $P_{\rm orb}$ and $T_{0}$ precisions for 23. The remaining 22 planets mostly corresponded to K2 targets revisited by TESS or ground-based facilities and still lacked photometric precision. We note that 11 of these 22 planets already had propagated $T_{0}$ uncertainties below the desired 5\,min threshold (Sect.~\ref{sec:ana_ephem}). Prior to our revisions, the best literature values for the 23 refined planets yielded a mean propagated $T_{0}$ uncertainty ($\sigma T_{0}$) of 65\,min. To date we have improved this number by 73$\%$, down to an average of 6\,min (Fig.~\ref{fig:improvement_hist}), and were able to reach better than 5\,min precision in 65$\%$ of the 23 planets.
In Fig.~\ref{fig:transits}, top panel, we show one of the NGTS transits obtained for ATREIDES, together with the most precise literature model (red dashed) and our updated model (solid green), showing a large timing offset of about two hours. In the bottom panel, we further show a simultaneous NGTS and SPECULOOS transit observation where we serendipitously detected a multi-transit event that allowed us to refine the ephemeris of two planets in the same system.  

Due to their interest for transit scheduling, we provide a catalog including the revised ATREIDES ephemeris (see Sect.~\ref{sec:discuss}). We will keep updating these values as new ground-based photometry and TESS data are acquired, incorporating these datasets in the dedicated planetary system characterizations that will be conducted in the upcoming ATREIDES publications. These analyses will be performed with more advanced transit pipelines (e.g., \citealt{Southworth2005,Southworth2011,Lendl2017,Lendl2020b,CastroGonzalez2023,Murgas2023,Chakraborty2024}), combining photometry and RVs whenever possible.


\subsection{Spectroscopy}
\label{sec:red_tr_spec}

ESPRESSO datasets are run through the \textsc{antaress} workflow (\citealt{Bourrier2024}), a set of methods to process high-resolution spectroscopic time-series in a robust way, extract accurate exoplanetary and stellar spectra, and analyze them for stellar, orbital, and atmospheric characterization. This ensures a homogeneous and reproducible analysis of the ATREIDES sample. We summarize here the standard procedures that we will use.   

\textsc{antaress} requires a number of assumed physical properties for the planetary system. They control the orbital and transit models, stellar line properties, and grid-based simulations of the full-disk and planet-occulted stellar surface. The workflow takes 2D echelle spectra as input, here extracted from ESPRESSO detector images, corrected for, and calibrated by the DRS pipeline (\citealt{Pepe2021}). Raw data are first set into \textsc{antaress} format and corrected for environmental and instrumental effects. The cleaned 2D spectra are then processed into the format relevant to the chosen analysis (e.g., conversion into 1D spectra or CCFs, binning between exposures and visits). Finally, various analysis can be applied to the stellar and planetary spectra in invididual or joint visits, to study the line shape or derive key system properties. We highlight that \textsc{antaress} was designed to avoid spectral resampling and propagate correlated noise, improving the quality of the processed data and reducing biases in their analysis.

The following corrections are typically applied to ESPRESSO data (see \citealt{Bourrier2024} for details):
\begin{enumerate}[leftmargin=*]
    \vspace{-0.5\baselineskip}
    \item Exposures, orders, and spectral ranges too noisy to be exploited are identified and removed. This screening is updated progressively, for example after identifying poorly-corrected telluric lines.
    \item Blaze profiles are used to retrieve the spectral flux calibration and detector noise profiles applied by the DRS. They are used by the workflow to scale back spectra into raw flux units and compute accurate weighted means.
    \item Telluric absorption lines, typically from H$_2$O and O$_2$, are automatically corrected with a model-based approach in individual exposures.
    \item Splines are used for normalization, correcting the low- and medium-frequency deviations of individual spectra from a reference stellar spectrum caused by Earth atmospheric diffusion.  
    \item Cosmics-affected pixels are automatically flagged through comparison with consecutive exposures, and replaced with their median.
    \item An analytical model of the complex interference pattern affecting ESPRESSO spectra is fitted over all exposures, allowing for a homogeneous correction without absorbing stellar or planetary signals.  
\end{enumerate}

Once corrected the ESPRESSO echelle spectra undergo the following processing steps:
\begin{enumerate}[leftmargin=*]
    \vspace{-0.5\baselineskip}
    \item Out-of-transit CCFs computed from the spectra are fitted with the adequate model to derive time-series of stellar line properties, used to identify outliers, characterize systematic trends, assess the relevance of using sky-corrected data, and measure the systemic RV in each epoch. Trends in the line shape and position are then directly corrected in the spectra.
    \item Spectra are Doppler-shifted into the star rest frame using the measured systemic RV and Keplerian model set by the input system properties. 
    \item Spectra are set to their true relative flux level, using light curves that can be analytical (for simple transits), \textsc{antaress}-based (for transits of multiple planets or across spotted and oblate stars), or imported (for more complex transits). Light curves can be chromatic to account for broabdand variations of the stellar intensity and planetary continuum.
\end{enumerate}

At this stage, echelle spectra undergo specific processing depending on the required analysis:
\begin{itemize}[leftmargin=*]
    \vspace{-0.5\baselineskip}
    \item Out-of-transit spectra are resampled over a common spectral grid, and binned together into a single master-out spectrum representative of the unocculted star. This 1D master is used to determine stellar bulk properties, such as rotational broadening or elementary abundances, and to build a weighted mask specific to the star for CCF computation.
    \item Spectra resampled over a common spectral grid are divided by the master-out stellar spectrum to compute 1D transmission spectra that can be fitted with numerical transit models, to constrain the planetary and stellar atmospheres (Sect.~\ref{sec:EVE}).
    \item Differential spectra are extracted between the master and each exposure. In-transit differential spectra are converted into intrinsic spectra tracing the specific intensity profile of planet-occulted regions. Intrinsic spectra, converted into CCFs, are analyzed with the RMR technique to characterize stellar line variations along the transit chord and to derive the planet spin-orbit angle (Sect.~\ref{sec:RMR}). Once the bulk photospheric properties are known, intrinsic spectra can be aligned and resampled over a common spectral grid, to be binned either into a high-S/N series function of the center-to-limb angle, or into a single master, allowing for the characterization of individual spectral lines from the spatially-resolved photosphere  (Sect.~\ref{sec:ana_star_bulk}). In parallel, out-of-transit differential spectra can be used to characterize stellar noise, in particular spot signatures from fast-rotating stars.
    \item Future versions of the \textsc{antaress} workflow will allow the absorption and emission lines from the planetary atmosphere, cleaned from stellar contamination, to be extracted from differential spectra.
\end{itemize}


\section{Interpretation plan}
\label{sec:ana_strategy}

To achieve our objectives (Sect.~\ref{sec:collab}) we will first study each ATREIDES system individually. Whenever relevant, we will revise the bulk stellar and planetary properties by analyzing the ATREIDES spectroscopic and photometric data together with published information. We will then determine the orbital architecture, internal structure, and atmospheric profile of the planets, and rewind their history by simulating their dynamical and atmospheric evolution together with that of the star. Once all ATREIDES systems have been characterized, we will perform statistical analyses of the full sample.  

\subsection{Bulk system properties}

\subsubsection{Star}
\label{sec:ana_star_bulk}

Stellar spectroscopic parameters ($T_{\mathrm{eff}}$, $\log g$, microturbulence, [Fe/H]) are estimated from the ESPRESSO spectra using the ARES+MOOG methodology (\citealt[][]{Sousa-21, Sousa-14, Santos-13}). Equivalent widths (EW) of stellar lines are consistently measured using the ARES code\footnote{The latest version, ARES v2, can be downloaded at \url{https://github.com/sousasag/ARES}} \citep{Sousa-07, Sousa-15}. Convergence for the best set of spectroscopic parameters is done by using a minimization process that finds the ionization and excitation equilibrium. This process makes use of a grid of Kurucz model atmospheres \citep{Kurucz1993} and the latest version of the radiative transfer code MOOG \citep{Sneden-73}. 

Individual stellar abundances of the elements are estimated with the classical curve-of-growth analysis method, by using the measured EWs of individual atomic lines in the spectrum. We assume local thermodynamic equilibrium and employ the same codes and models that are used for the stellar parameter determinations. For the derivation of chemical abundances of refractory elements we closely follow the methods described in, e.g.,  \citet{Adibekyan-12, Adibekyan-15,Delgado2017}. Abundances of the volatile elements, C and O, are derived following the method of \cite{Delgado2021,Bertrandelis2015}. In the specific case of oxygen, a careful visual analysis is performed around the weak forbidden oxygen line at 6300.3\,\AA\, to remove spectra contaminated by tellurics or an oxygen airglow at the same wavelength, which can lead to erroneous oxygen abundance measurements. All the [X/H] ratios are obtained by doing a differential analysis with respect to a high S/N solar spectrum. 

Our spectroscopic analyses will be primarily performed on a high-S/N master spectrum of the disk-integrated star, and is most efficient for F, G, and K-type stars. For A- and M-type stars, or fast-rotating stars, we will use different methodologies \citep[][]{AntoniadisKarnavas2020, Tsantaki2018}. In particular we will also perform our analyses on the master spectrum extracted from the ESPRESSO data along the transit chord (Sect.~\ref{sec:red_tr_spec}), which probes the local photosphere and is much less affected by rotational broadening.  

When relevant, we will use SED fitting and evolutionary stellar models, informed by the retrieved spectroscopic parameters and GAIA parallax measurements, to revise fundamental stellar parameters (radius, mass, and ages).


\subsubsection{Planet}
\label{sec:ana_pl_bulk}

When relevant the transit photometry on our targets is used to refine their radius. Planet masses are taken from the literature, as they are unlikely to be improved by ATREIDES ESPRESSO data acquired over short durations. Transit photometry in multi-planet systems may, however, refine planet masses through TTV characterization (e.g., \citealt{Leleu2024}).

Mass and radius are used to determine the planet internal structure, using the \textsc{jade} code (\citealt{Attia2021}; Attia et al., in prep) and the \textsc{plaNETic} framework (\citealt{Egger2024}), based on the \textsc{biceps} internal structure model (\citealt{Haldemann2024}). Chemical abundances of the host star photosphere (Sect.~\ref{sec:ana_star_bulk}) may provide further constraints on the planet interior composition (e.g., \citealt{Dorn2017a,Thiabaud2015}), although this approach should be taken with caution (e.g., \citealt{Adibekyan2011})

The \textsc{jade} planetary structure currently consists in an iron core ($\alpha$-Fe), a silicate mantle (\ce{MgSiO3}), and an H/He-dominated envelope. The latter is divided into an upper region absorbing the stellar radiation and a lower region redistributing energy via radiation and convection. The Rosseland mean opacity of the envelope can be increased by including trace amount of metals, assuming Solar abundances. Starting from the top, the code integrates the 1D thermodynamical structure of the envelope and the polytropic equation of state (\citealt{Seager2007}) of the core and mantle, iterating over a grid of radii until finding the one that yields zero mass at planet center (e.g., \citealt{Lopez2013,Jin2014}). \textsc{jade} fits for the mantle-to-planet mass fraction, envelope-to-planet mass fraction, and envelope metallicity with a MCMC approach, using the measured planet radius as constraint. The planet mass, orbital properties, and atmospheric He abundance are fixed. The retrieval is performed at a given age for the system, to account for the impact of the stellar irradiation and planetary internal luminosity on the atmospheric structure. The stellar luminosity is set to evolutionary curve models constrained by the star's current-day properties (Sect.~\ref{sec:ev_sim_star}). This simplified approach is driven by the intended use of \textsc{jade}, which is to simulate the secular atmospheric and orbital evolution of the planet for a given primordial mass and orbit (Sect.~\ref{sec:ev_sim_pl}). The core+mantle mass and envelope metallicity remain constant over time in a \textsc{jade} simulation, so that the primordial internal structure can be self-consistently set to the derived present-day values.


Alternatively, we follow the three-steps approach of the \textsc{plaNETic} framework\footnote{\url{https://github.com/joannegger/plaNETic}}. First, we sample relevant model parameters and solve the required differential equations to compute an internal structure database. We then train a neural network to emulate this first step, providing a much faster way to compute a planetary radius as a function of its properties. Finally, we use this network to sample the PDF of the model parameters, fitting for the measured transit depth and RV semi-amplitude. Internal structure models are computed with the \textsc{biceps} code (\citealt{Haldemann2024}) as a function of their mass, composition (mass fraction of core, mantle, volatile layer), age, orbital separation. The planet is made of a core (iron and \ce{FeS}), a mantle (\ce{SiO2},\ce{MgO}, and \ce{FeO}) and a volatile layer (mixture of H/He and water). We typically explore two possibilities regarding the volatile layer. The wet case assumes formation beyond the iceline, resulting in water being accreted as ice and forming a uniformly mixed volatile layer together with the H/He after inward migration. The core, mantle, and water mass fractions (relative to the non H/He part of the planet) are sampled uniformly, enforcing their sum to be one and setting an upper limit of 0.5 for water (\citealt{Thiabaud2014,Marboeuf2014}). The H/He mass fraction (relative to the whole planet) is sampled from a uniform-in-log distribution. All four mass fractions are then re-scaled to add-up to one. The dry case assumes formation inside the iceline, resulting in water and H/He being accreted together and mixed in a common layer. The water fraction in the gas is assumed to be a Gaussian with mean 0.5 \% and standard deviation 0.25 \% (\citealt{Mousis2009}). The core and mantle are modeled as in the wet case.


\subsection{Orbital architecture}
\label{sec:RMR}

Intrinsic CCFs, which trace photospheric properties along the transit chord, are derived from ESPRESSO spectra (Sect.~\ref{sec:red_tr_spec}) and analyzed with \textsc{antaress} following the RMR procedure (\citealt{Bourrier2021,Bourrier2024}).

In a first step, intrinsic CCFs are fitted independently to determine which exposures to include in the joint fit, and evaluate the best models describing the photospheric RV field, the local stellar line profile, and its spatial variations across the photosphere. We typically discard limb CCFs because of their low S/N, as well as CCFs with properties strongly deviating from the expected series, e.g. due to stellar activity. The photospheric RV model can describe solid-body (SB) or differential (DR) rotation, and account for center-to-limb convective blueshift variations (CB). Parameters controlling the line shape are described as polynomials of a chosen photospheric coordinate, typically the sky-projected distance to star center. Priors on the line model parameters are informed by the stellar bulk properties and disk-integrated line.  

In a second step, a stellar line model series informed by the first step is fitted to all intrinsic CCFs simultaneously, exploiting their full spectro-temporal information. Multiple planets' transits, and if necessary datasets from different instruments, can be fitted together, bringing strong constraints on shared properties like the stellar rotational velocity. The model defines the local line before instrumental convolution, so that the derived photospheric properties are comparable between spectrographs. 

Unless DR can be constrained, RMR fits to transit spectroscopy alone yield the sky-projected spin-orbit angle $\lambda$ and stellar rotational velocity $v_{\rm eq} \sin i_\star$. If the equatorial rotation period $P_\mathrm{eq}$ can be estimated from the modulation of stellar activity indexes or photometry (Sect.~\ref{sec:obs_photom}), or if the stellar inclination $i_{\star}$ can be measured from asteroseismology or spotted light curves, the RMR fit is performed with the independent variables $R_\star$, $P_\mathrm{eq}$, and $\cos i_\star$ (\citealt{Masuda2020}). The known or derived PDF for $i_{\star}$ can then be combined with those of $\lambda$ and the orbital inclination $i_\mathrm{p}$ to sample the 3D spin-orbit angle $\Psi$ (e.g., \citealt{Bourrier2023}). There generally remains a degeneracy between equiprobable ``northern'' ($\Psi^\mathrm{N}$, $i_\star$) and ``southern'' ($\Psi^\mathrm{S}$, $\pi - i_\star$) configurations, which can be combined when similar. If the spin-orbit angles of two or more planets in the system can be measured, it is possible to derive the mutual inclination between their orbital planes (e.g., see Sect. 5 in \citealt{Bourrier2021}).

The RMR technique and its \textsc{antaress} implementation were recently upgraded to incorporate spots on the stellar surface (Mercier et al., in prep.). ESPRESSO transits contaminated by spot-crossing events and/or fast-rotating spots on young stars will be interpreted with this approach. 


\subsection{Upper planetary atmosphere}

The ATREIDES datasets yield time-series of transmission spectra (Sect.~\ref{sec:red_tr_spec}) that can be interpreted in terms of atmospheric composition and dynamics using numerical models of the stellar and planetary atmospheres.

\subsubsection{Probing Neptunian atmospheres}

Much of the processes governing the atmospheric structure and chemistry of Neptunian worlds remain unknown. Detections of massive atmospheric outflows from hot Neptunes (e.g., \citealt{Ehrenreich2015}) changed the paradigm regarding these worlds' evolution (e.g., \citealt{Owen2018}). Hydrodynamical escape may remove a substantial fraction of Neptunian atmospheres (e.g., \citealt{Bourrier2018_GJ3470b}), in particular their H/He content, leading to the formation of metal-rich secondary atmospheres (e.g., \citealt{Hu2015}).   

The outermost layers, thermosphere and exosphere, can be probed via hydrogen and helium. Helium, routinely used to probe even Neptune-size planets (\citealt{Allart2018, Zhang2021, Zhang2023, Allart2023, Zhang2024}), does not have accessible transitions within the ESPRESSO optical range. Hydrogen, historically probed from space in the UV Lyman-$\alpha$ line (e.g., \citealt{VM2003,Lecav2012}), has multiple optical transitions from the Balmer series. Depending on the high-energy irradiation from the host star, the most prominent Balmer transition (H-$\alpha$, \citealt{Orell2024}) might be the optimal tracer of atmospheric escape in the ATREIDES sample. We note that NLTE effects may have to be considered when modelling the atmosphere of close-in Neptunes, especially with a tracer like H-$\alpha$ (e.g., \citealt{Kubyshkina2024}).

Some planets along the Desert rim have shown absorption signatures comparable in amplitude to the larger hot Jupiters, e.g. in the sodium lines of WASP-166~b \citep{Seidel2020b, Seidel2022} and WASP-127~b (which also displayed potassium and lithium absorption, \citealt{Chen2018, Allart2020}). The sodium doublet warrants particular attention because it is an effective tracer of wind dynamics in the stratosphere, up to the thermosphere for hot and ultra-hot Jupiters \citep{Seidel2020a, Seidel2021, Mounzer2022, Seidel2023a}. Further detections of sodium across the Desert and Savanna could open pathways to study the irradiation threshold and altitudes above which Neptunian atmospheres lose their stability and undergo hydrodynamical escape.

At lower altitudes within the stratosphere, hot and ultra-hot Jupiters have shown absorption signatures from the lines of ions \citet{Borsa2021,Azevedo2022} and alkali metals \citet{Borsa2021, Damasceno2024}. The hottest planets in the ATREIDES sample represent exciting candidates to extend this window to Neptunian worlds, and assess their possible metal enrichment.

\subsubsection{Interpreting transmission spectra}
\label{sec:EVE}

Transmission spectra are typically calculated using the unocculted disk-integrated stellar spectrum as proxy. The dissimilarity between the disk-integrated and planet-occulted stellar lines creates distortions (POLDs, \citealt{dethier2023,Carteret2023}) that blend with the planetary signatures. While it is common practice, correcting transmission spectra for these distortions inherently introduces biases (\citealt{Dethier2024}). A proper interpretation requires numerical models able to reproduce the complete distorted transmission spectra. 

We use the Evaporating Exoplanets (\textsc{EvE}) code \citep{bourrier2013,bourrier2015,Bourrier2016}, which simulates stellar spectra through the occultation of a spatially- and spectrally-resolved stellar disk by a transiting planet, accounting for its 3D orbital architecture and atmospheric structure. Transit analysis with \textsc{antaress} constrains the orbital architecture and stellar surface motions (Sect.~\ref{sec:RMR}), while providing intrinsic stellar spectra (Sect.~\ref{sec:red_tr_spec}) that can directly tile the stellar grid or constrain theoretical stellar spectra. As such, \textsc{EvE} accurately reproduces the center-to-limb variations (CLV) in stellar lines absorbed by the planetary atmosphere, responsible for the POLDS (\citealt{dethier2023}). This approach also allows us to incorporate active regions, which may be relevant for variable stellar lines like H-$\alpha$ \citep{Huang2023}. \textsc{EvE} delivers realistic high-resolution, high-cadence disk-integrated stellar spectra, which can then be processed to match the resolution and cadence of a given spectrograph. This self-consistent approach eliminates the need for corrections, reducing the risk of biases when comparing synthetic and observed transmission spectra.


\subsection{Secular evolution}

\subsubsection{Star}
\label{sec:ev_sim_star}

The Geneva stellar evolution code \citep[\textsc{GENEC},][]{Eggenberger2008} is used to compute rotating models of the host star. These models follow the evolution of the star accounting for the effects of hydrodynamic and magnetic instabilities \citep{Eggenberger2022}. This enables us to study the impact of the rotational history of the host star, which is unknown, on the global evolution of the planetary system, by computing models representative of slow, medium and fast rotating stars observed in open clusters \citep{Eggenberger2019}. The high energy fluxes emitted by the host star are obtained from the structural, rotational and magnetic properties of the models. Each rotational history then corresponds to a different evolution of the X-ray luminosity of the host star and to the corresponding extreme ultraviolet luminosity \citep[see e.g.][]{Pezzotti2021}. The present models use the relation between the X-ray luminosity and the EUV luminosity as given by \citet{Sanz-Forcada2011}. This leads to slightly higher L$_\mathrm{EUV}$, and hence total L$_\mathrm{XUV}$, luminosities than the ones that would be obtained by using the relation between the surface X-ray and EUV fluxes as given by \citet{Johnstone2021} or the recent revised relation between the X-ray and EUV luminosities reported by \citet{SanzForcada2025} (typical difference of about 25\% in L$_\mathrm{EUV}$ with \citealt{Sanz-Forcada2011} at the estimated age of the system). In addition to the evolution of the bolometric and high energy fluxes of the host star, following its rotational properties enables to investigate the exchange of angular momentum between the star and the orbit of the planet. The role of tides in shaping the architecture of the system can be studied by accounting for dissipation by equilibrium tides and dynamical tides through inertial waves in the convective envelope of the host star \citep[][]{Rao2018}.

\subsubsection{Planet}
\label{sec:ev_sim_pl}

The \textsc{jade} code (\citealt{Attia2021}, Attia et al. in prep.)\footnote{We used version 1.0.0 of the code, publicly available at \url{https://gitlab.unige.ch/spice_dune/jade}.} was designed primarily to investigate the late HEM and delayed evaporation of a close-in giant planet, following its trapping into Kozai-Lidov resonance by a massive outer companion. It averages the equations of motion over the bodies' orbits to simulate the dynamical evolution of the system over long timescales while simultaneously integrating the inner planet atmospheric structure. This sets \textsc{jade} apart from N-body integrators that simulate dynamical evolution accurately over short timescales with no or basic planetary structure (e.g., \citealt{Beust2012,Benbakoura2019}), from detailed models of escaping atmospheres fixed in time or applying simple dynamical evolution (e.g., \citealt{Salz2016b,Kubyshkina2018}), and from population syntheses that necessarily use idealized dynamical and atmospheric prescriptions (e.g., \citealt{Kurokawa2014,Jin2018}). 

A simulation is initialized at a chosen age, with a given orbit and atmospheric mass fraction for the inner planet. Its core, mantle mass, and atmospheric abundance are fixed and determined through an internal structure retrieval informed by the present-day properties (Sect.~\ref{sec:ana_pl_bulk}). The orbital properties and mass of the outer companion are fixed. The stellar bolometric and XUV luminosity curves, which impact the evolution of the planetary atmosphere, are set to the outputs of stellar evolution codes like \textsc{genec} (Sect.~\ref{sec:ev_sim_star}). \textsc{jade} updates the planetary radius and mass over time, as variations in stellar irradiation (due to changes in the orbit and intrinsic stellar luminosity) and core-powered heating control the temperature of the atmosphere and its erosion. Mass loss is calculated using the analytical formulae from \citet{Salz2016b}. In parallel, the code integrates the evolution of the planetary orbital properties over the relevant dynamical timescales, accounting for changes in gravitational interactions with the star and companion due to the evolving planetary structure. If relevant, the evolution of the planetary atmosphere can be simulated with fixed orbital properties (e.g., assuming early disk-driven migration to a stable orbit), or the evolution of the planetary orbit can be simulated with a fixed atmospheric structure (e.g., assuming the planet is always faintly irradiated by its host star). 

\textsc{jade} evolutionary simulations of a planetary system, constrained by measurements of its orbital architecture, stellar and planetary properties, aim at determining the primordial orbit and envelope mass of the inner planet, and the orbit and mass of the putative outer companion. Given \textsc{jade} computing time we use importance sampling to explore the parameter space. A grid of simulations spanning the full life of the system is run to coarsely sample the PDFs of the fitted properties. The probability of a given simulation is derived from the integral over all timesteps of the log-likelihood for the measured properties with respect to the model, accounting for the relevant prior distributions. The interpolated probability grid from the simulations is then used to draw samples uniformly over the parameter space and build their final PDF.


\section{First ATREIDES result: the TOI-421 system}
\label{sec:TOI421}

The TOI-421 is a bright (V = 9.9) G9-dwarf star known to host two short-period transiting planets (\citealt{Carleo2020}). The warm Neptune TOI-421 c, located within the Savanna, was the first planet observed as part of ATREIDES on 2023 Nov. 06. The sub-Neptune TOI-421 b, on a shorter period, was previously observed as part of the ESPRESSO GTO on 2022 Nov. 17. We analyzed the two datasets to benefit from their joint analysis in better understanding the overall architecture of the system. Planetary atmospheres will be studied in a follow-up paper.

\begin{table*}  
\caption[]{Properties of the TOI-421 system.}
\centering
\begin{threeparttable}
\begin{tabular}{c c c c c}
\hline
\hline 

Parameter & Symbol & Value  &  Unit   &   Origin  \\

\hline

\multicolumn{5}{c}{\textit{Stellar parameters}} \\ 
Spectral type       &               & G9 V & & \citet{Carleo2020} \\
Temperature & $T_{\rm eff}$ & 5291$\pm$64 & K &  \citet{Krenn2024} \\
Surface gravity  & log\,$g$ & 4.48$\pm$0.03 &  &  \citet{Krenn2024} \\ 
Radius      & $R_\star$     & 0.866$\pm$0.006 & $R_\odot$ &  \citet{Krenn2024} \\
Mass      & $M_\star$     & 0.833$\stackrel{+0.048}{_{-0.054}}$ & $M_\odot$ &  \citet{Krenn2024} \\
Spin inclination & $i_{\star \rm}^\mathrm{N}$ & 46.9$\stackrel{+4.2}{_{-4.5}}$ & deg & This work \\
                         & $i_{\star \rm}^\mathrm{S}$ & 133.1$\stackrel{+4.5}{_{-4.2}}$ & deg & This work \\
Equatorial period & $P_{\rm eq}$ & 19.8$\pm$0.8 & d & This work \\
Projected velocity & $v_\mathrm{eq} \sin i_\star$ & 1.60$\pm$0.09 & km s$^{-1}$ & This work \\
Age & $\tau$ & 10.9$^{+2.9}_{-5.2}$ & Gyr & \citet{Krenn2024} \\
Limb-darkening coefficients & $u_1$ & 0.48$\pm$0.02 & &  \citet{Krenn2024}  \\
                            & $u_2$ & 0.18$\pm$0.02	 & &  \citet{Krenn2024} \\
\hline
\multicolumn{5}{c}{\textit{Planet b parameters}} \\ 
\hline
Orbital period & $P$ & 5.197576$\pm$0.000005 & d & \citet{Krenn2024} \\
Transit epoch & $T_0$ & 2459189.7341$\pm$0.0005  & BJD$_{\rm TDB}$ & \citet{Krenn2024} \\
Eccentricity & $e$ & 0.13$\pm$0.05 & & \citet{Krenn2024} \\
Argument of periastron & $\omega$ & 140$\pm$30 & deg & \citet{Krenn2024} \\
Stellar reflex velocity & $K$ & 2.83$\pm$0.18 & m s$^{-1}$ & \citet{Krenn2024} \\
Planet mass & $M_{\rm p}$ & 6.7$\pm$0.6     & $M_\Earth$ & \citet{Krenn2024} \\
Scaled separation & $a/R_\star$ & 13.76$_{-0.22}^{+0.23}$ & & \citet{Krenn2024} \\
Orbital inclination & $i$ & 85.67$^{+0.30}_{-0.21}$ & deg & \citet{Krenn2024} \\
Impact parameter & $b$ & 0.942$\pm$0.011 & & \citet{Krenn2024} \\
Transit duration & $T_{14}$ & 1.090$\pm$0.034 & h & \citet{Krenn2024} \\
Planet-to-star radius ratio & $R_{\rm p}/R_\star$ & 0.0279$\pm$0.0008     & & \citet{Krenn2024} \\
Planet radius & $R_{\rm p}$ & 2.64$\pm$0.08     & $R_\Earth$ & \citet{Krenn2024} \\
Projected spin--orbit angle & $\lambda_\mathrm{b}$ & -35$\stackrel{+26}{_{-23}}$ & deg & This work \\
3D spin--orbit angle & $\psi^{\rm N}_\mathrm{b}$ & 54$\stackrel{+12}{_{-16}}$ & deg & This work \\
                     & $\psi^{\rm S}_\mathrm{b}$ & 61$\stackrel{+11}{_{-14}}$  & deg &  \\
                     & $\psi_\mathrm{b}$ & 57$\stackrel{+11}{_{-15}}$  & deg &  \\
 
\hline
\multicolumn{5}{c}{\textit{Planet c parameters}} \\ 
\hline
Orbital period & $P$ & 16.067541$\pm$0.000004 & d & \citet{Krenn2024} \\
Transit epoch & $T_0$ & 2459195.30741$\pm$0.00018 & BJD$_{\rm TDB}$ & \citet{Krenn2024} \\
Eccentricity & $e$ & 0.19$\pm$0.04 & & \citet{Krenn2024} \\
Argument of periastron & $\omega$ & 102$\pm$14 & deg & \citet{Krenn2024} \\
Stellar reflex velocity & $K$ & 4.1$\pm$0.3    & m s$^{-1}$ & \citet{Krenn2024} \\
Planet mass & $M_{\rm p}$ & 14.1$\pm$1.4     & $M_\Earth$ & \citet{Krenn2024} \\
Scaled separation & $a/R_\star$ & 29.0546178$\pm$0.40 & & \citet{Krenn2024} \\
Orbital inclination & $i$ & 88.30$\pm$0.05  & deg & \citet{Krenn2024} \\
Impact parameter & $b$ & 0.70$\pm$0.03    & & \citet{Krenn2024} \\
Transit duration & $T_{14}$ & 2.755$\pm$0.022 & h & \citet{Krenn2024} \\
Planet-to-star radius ratio & $R_{\rm p}/R_\star$ & 0.0540$\pm$0.0006   & & \citet{Krenn2024} \\
Planet radius & $R_{\rm p}$ & 5.09$\pm$0.07     & $R_\Earth$ & \citet{Krenn2024} \\
Projected spin--orbit angle & $\lambda_\mathrm{c}$ & 14.0$\pm$1.8 & deg & This work \\
3D spin--orbit angle & $\psi^{\rm N}_\mathrm{c}$ & 44.5$\stackrel{+4.8}{_{-4.4}}$ & deg & This work \\
                     & $\psi^{\rm S}_\mathrm{c}$ & 45.3$\stackrel{+4.7}{_{-4.4}}$  & deg &  \\
                     & $\psi_\mathrm{c}$ & 44.9$\stackrel{+4.4}{_{-4.1}}$  & deg &  \\
	    
\hline
\end{tabular}
\begin{tablenotes}[para,flushleft]
\end{tablenotes}
\end{threeparttable} 
\label{tab:TOI421_prop}
\end{table*}

\newpage


\subsection{Observations and data reduction}

\begin{figure}
\includegraphics[trim=0cm 0cm 0cm 0cm,clip=true,width=\columnwidth]{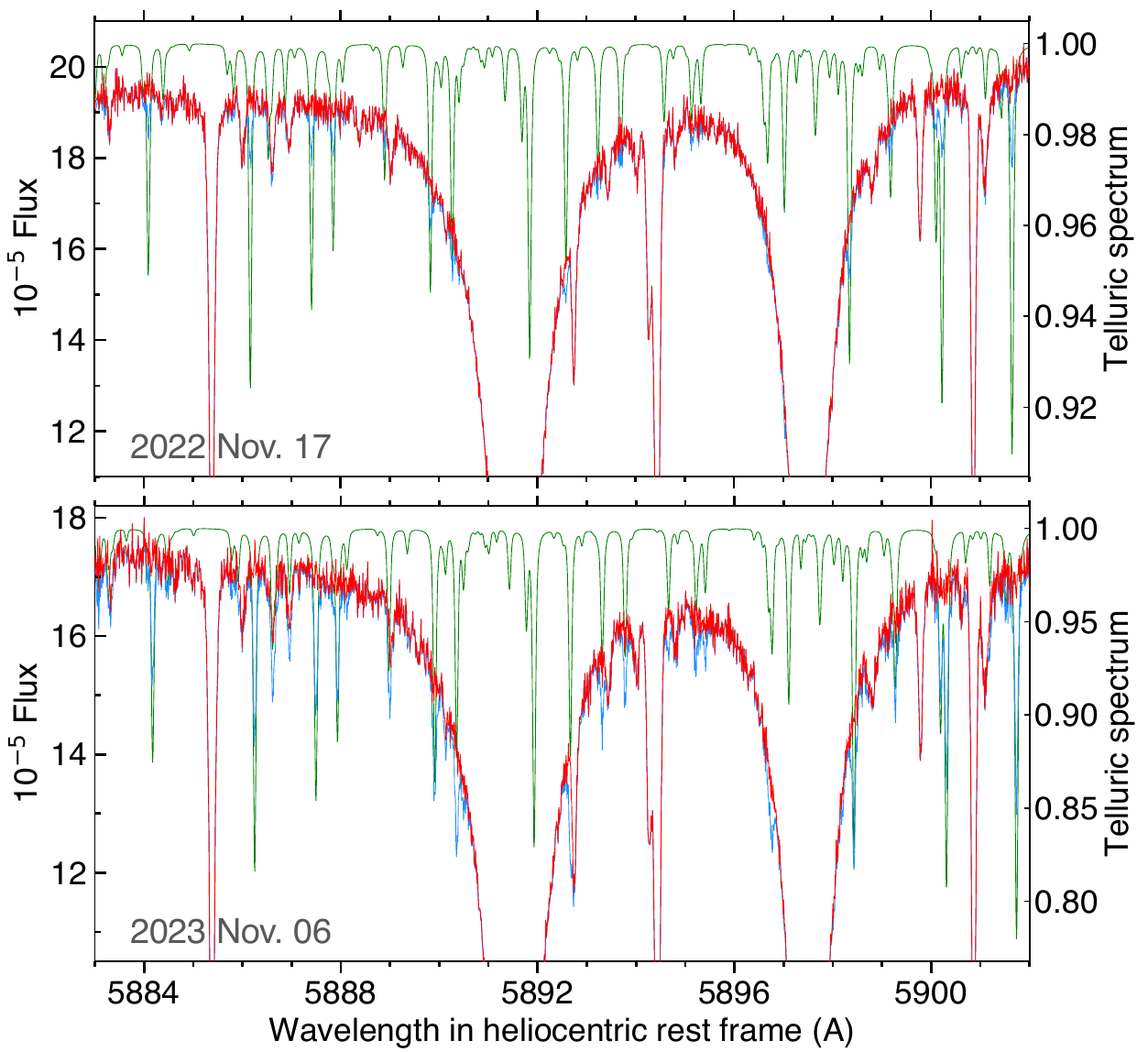}
\centering
\caption[]{Mid-visit ESPRESSO spectra of TOI-421, illustrating telluric correction in the sodium doublet. Spectra before/after correction are shown in red/blue, respectively, with the best-fit telluric model in green.} 
\label{fig:tell}
\end{figure}

\begin{figure*}
\begin{minipage}[tbh!]{\textwidth}
\includegraphics[trim=0cm 0cm 0cm 0cm,clip=true,width=\textwidth]{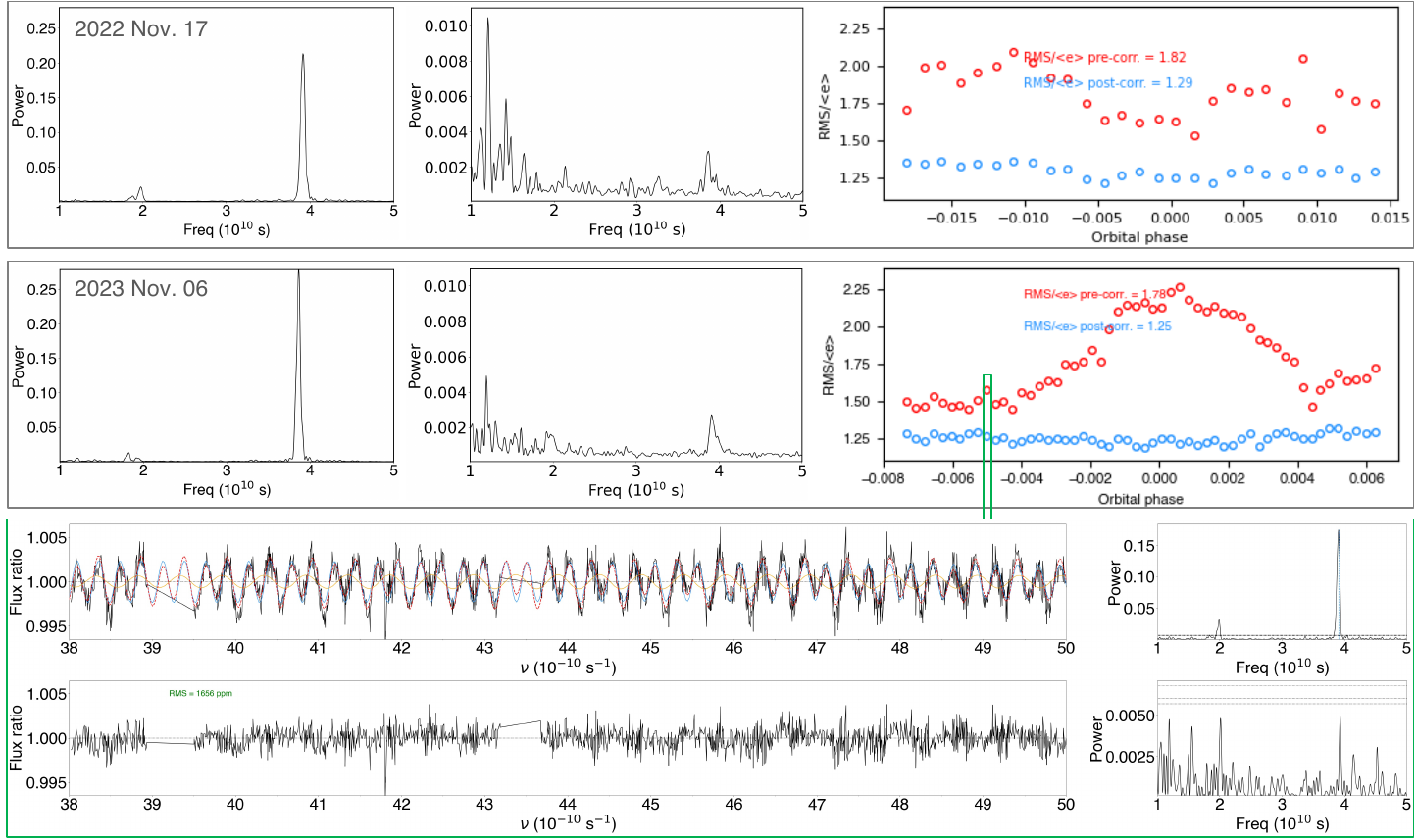}
\centering
\end{minipage}
\caption[]{ESPRESSO wiggle correction. \textit{Top panels}: Periodogram cumulated over all transmission spectra, before (left) and after (middle) correction. Right panels shows the ratio between the RMS of the transmission spectra and their median error before (red) and after (blue) correction.  \textit{Bottom panel}: Example transmission spectrum as a function of light frequency, before (top, overplotted with the best-fit wiggle model in red with its dominant blue and secondary orange components) and after (bottom) correction.}
\label{fig:wiggle}
\end{figure*}

ESPRESSO data, reduced with version 3.0.0 of the DRS pipeline, were processed and analyzed with the \textsc{antaress} workflow (Sect.~\ref{sec:red_tr_spec}). All exposures were obtained with high quality (S/N between 72--128), except for the last exposure on 2023 Nov. 06 which shows outlying stellar line properties and was excluded. Telluric absorption was corrected for H$_{2}$O and O$_{2}$ (Fig.~\ref{fig:tell}), showing well-behaved fitted atmospheric properties. Flux balance analysis revealed interference patterns at medium frequencies in 2022 Nov. 17 similar to those observed in other datasets (\citealt{Mounzer2025}), which were corrected for using fine-smoothing splines
. Correction of the standard wiggle beat pattern (\citealt{Bourrier2024}) then removed most of the remaining noise structure (Fig.~\ref{fig:wiggle}). We accounted for an unexplained reset in wiggle properties (mimicking a guide star change) at about mid-transit in 2023 Nov. 6, and for an effective guide star change in 2022 Nov. 17.

\begin{figure}
\includegraphics[trim=0cm 0cm 0cm 0cm,clip=true,width=\columnwidth]{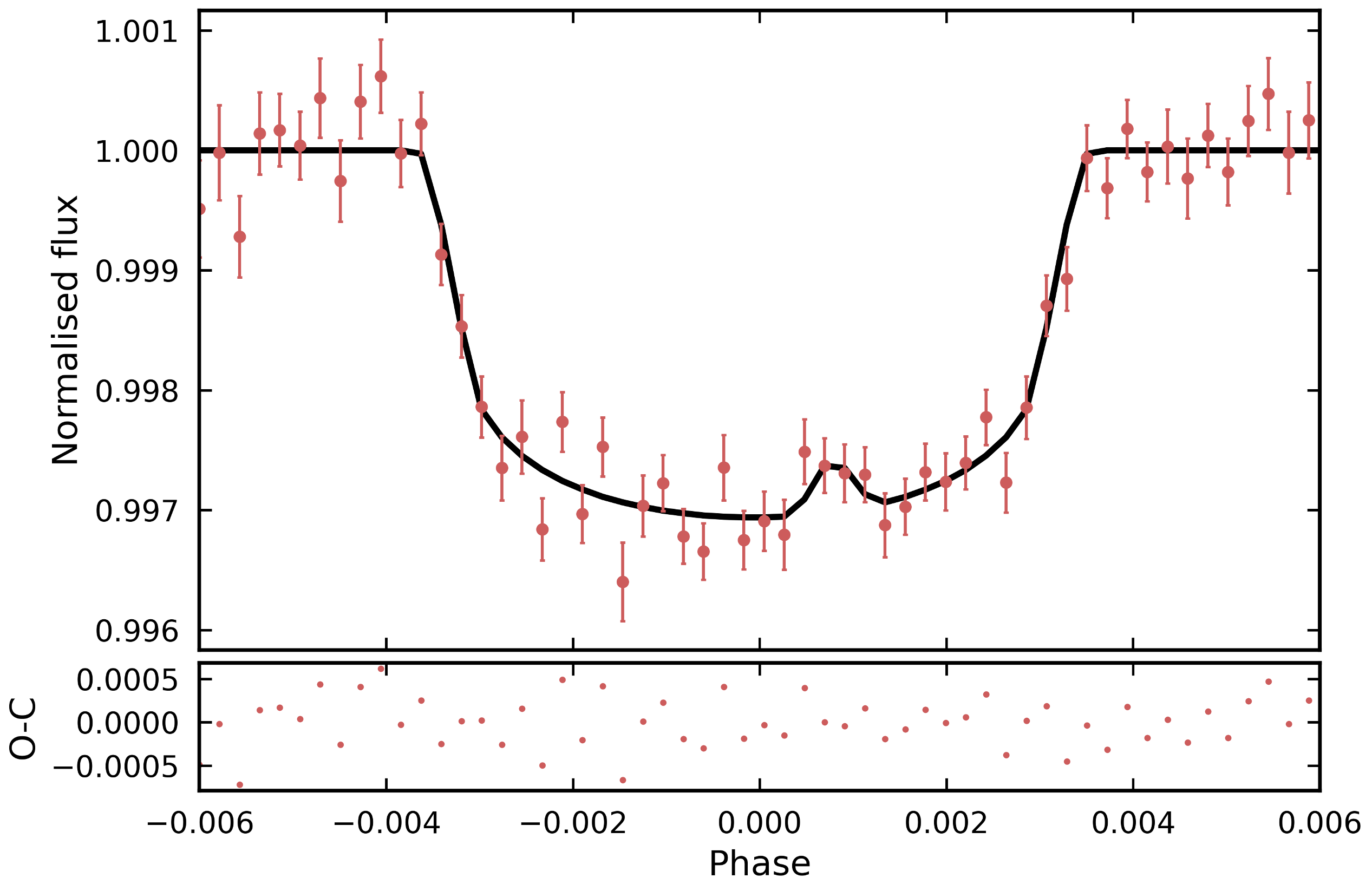}
\centering
\caption[]{NGTS transit light curve of TOI-421~c, at the nominal cadence (red points). The black curve is the best-fit \textsc{sage} model accounting for the spot-crossing.} 
\label{fig:NGTS_LC}
\end{figure}

With no significant TTV detected in the system, the ephemeris of both TOI-421 planets are known to a sufficient precision from \citet{Krenn2024} that photometric follow-up was not required. Nonetheless we used the NGTS telescopes to observe the full photometric transit of TOI-421~c on 2023 Nov. 06, simultaneously with our ESPRESSO observation. We used a combination of ten cameras and a 10\,s exposure time. Light curves were extracted as described in Sect.~\ref{sec:obs_dedicated_photometry}, revealing a possible spot-crossing by the planet (Fig.~\ref{fig:NGTS_LC}). 

\subsection{Revision of system properties}

Most of the system properties were taken from the comprehensive study performed by \citealt{Krenn2024} (Table \ref{tab:TOI421_prop}). As a comparison we re-evaluated the stellar spectroscopic parameters from our ESPRESSO data (Sect.~\ref{sec:ana_star_bulk}). We used the list of iron lines presented in \citet[][]{Sousa2008}, and measured a trigonometric surface gravity (using recent GAIA data, \citealt{Sousa-21}) that is consistent but more accurate than the spectroscopic surface gravity. Our derived stellar parameters are fully consistent with \citet[][]{Krenn2024}. Spectral lines are broadened by stellar rotation ($v_{\rm eq} \sin i_\star$), micro- ($V_\mathrm{mic}$) and macro- ($V_\mathrm{mac}$) turbulences, and the spectrograph response. Our analysis of the master-out ESPRESSO stellar spectrum (Sect.~\ref{sec:ana_star_bulk}) yields a total line broadening of 3.27$\pm$0.02\,km\,s$^{-1}$. By subtracting quadratically the rotational broadening derived from our RM analysis (1.60$\pm$0.09\,km\,s$^{-1}$, Sect.~\ref{sec:orb_TOI421}) we deduce a combined $V_\mathrm{mic}$ and $V_\mathrm{mac}$ broadening of 2.83$\pm$0.04 \,km\,s$^{-1}$. This is perfectly in line with the 2.86\,km\,s$^{-1}$ derived from calibration relations ($V_\mathrm{mic}$ = 0.90\,km\,s$^{-1}$, \citealt{Bruntt2010} ; $V_\mathrm{mac}$ = 2.71\,km\,s$^{-1}$, \citealt{Doyle2014}).

We jointly modeled the NGTS transit light curves of TOI-421~c from all cameras (Fig.~\ref{fig:NGTS_LC}) using \textsc{conan} (COde for exoplaNet ANalysis, \citealt{Lendl2017}). To account for correlated noise arising from instrumental systematics, weather, or stellar activity each light curve was fitted with a specific photometric baseline model together with the common transit model. We tested different sets of parametrization for the baseline, including a combination of time, shifts of stellar point spread function, airmass and the peak target flux. The \cite{Krenn2024} values were used to fix system parameters and set normal priors on the transit duration, mid-transit time, and impact parameter. A wide uniform prior $\mathcal{U}$(0, 0.1) was set on R$_{p}$/R$_{\star}$. Limb-darkening was modeled using a quadratic law, with normal priors on the coefficients centered on estimates from LDCU\footnote{https://github.com/delinea/LDCU}. The transit model accounts for the presence of a spot, simulated with the \textsc{sage} code (\citealt{Chakraborty2024}) for a given position, size and temperature.

We first simulated the internal structure of TOI-421~c with the \textsc{jade} code (Sect.~\ref{sec:ana_pl_bulk}). The stellar luminosity was set to the GENEC evolutionary curve (Sect.~\ref{sec:evol_TOI421}) at an age of 10.9\,Gyr. The mass and orbital properties of the planet were set to their nominal present-day values (Table~\ref{tab:TOI421_prop}) and the He mass fraction of gaseous envelope to Neptune's value (0.2). Uniform priors ($\mathcal{U}$(0,1)) were set on the atmosphere and mantle mass fractions, and a narrower prior ($\mathcal{U}$(0,0.1)) on the trace atmospheric metallicity. The retrieval is constrained by the measured planet radius. The correlation diagram for the fitted and derived properties is shown in Fig.~\ref{fig:Corr_diag_JADE_int_PAPER}. Atmospheric metallicity, and the repartition of the core and mantle mass, are not constrained. We derive an envelope mass $\log(M_\mathrm{Env} [M_\oplus])$ = 0.285$\stackrel{+0.057}{_{-0.085}}$, which represents a mass fraction $f_\mathrm{Env/Pl}$ = 0.137$\stackrel{+0.016}{_{-0.027}}$ relative to the whole planet. This is broadly consistent with the internal structure retrieval by \citealt{Krenn2024} ($\log(M_\mathrm{Env} [M_\oplus])$ = 0.34$\stackrel{+0.12}{_{-0.16}}$), and with the mass fractions they derived from evolutionary models ($f_\mathrm{Env/Pl}$ = 0.106$\pm$0.003 and 0.242$\pm$0.014). 

We then simulated the internal structures of TOI-421~b and c with the \textsc{biceps} code (Sect.~\ref{sec:ana_pl_bulk}), with results shown in Appendix~\ref{apn:corr_instr}. We conclude that the water mass fraction is essentially unconstrained for both planets in the wet case, with a slight preference for low mass fraction in planet c. The core mass fraction is relatively small (mean of 12 \%) for both planets, while the gas fraction is smaller for planet b (mean $\log fm_{\rm gas} = -1.45$) than planet c (mean $\log fm_{\rm gas} = -0.65$). In the dry case the core mass fraction is slightly larger (around 16 \% for both planets), and the gas mass fraction is slightly decreased (mean $\log fm_{\rm gas}$ of -2.03 and -0.70 respectively for the two planets), the mass fraction of the mantle increasing substantially to 84 \% (mean value) for both planets. The dry case for TOI-421~c, which is the most comparable to the \textsc{jade} retrieval, corresponds to an envelope mass fraction of 0.20.


\subsection{Orbital architecture}
\label{sec:orb_TOI421}

\begin{figure}
\includegraphics[trim=0cm 0cm 0cm 0cm,clip=true,width=\columnwidth]{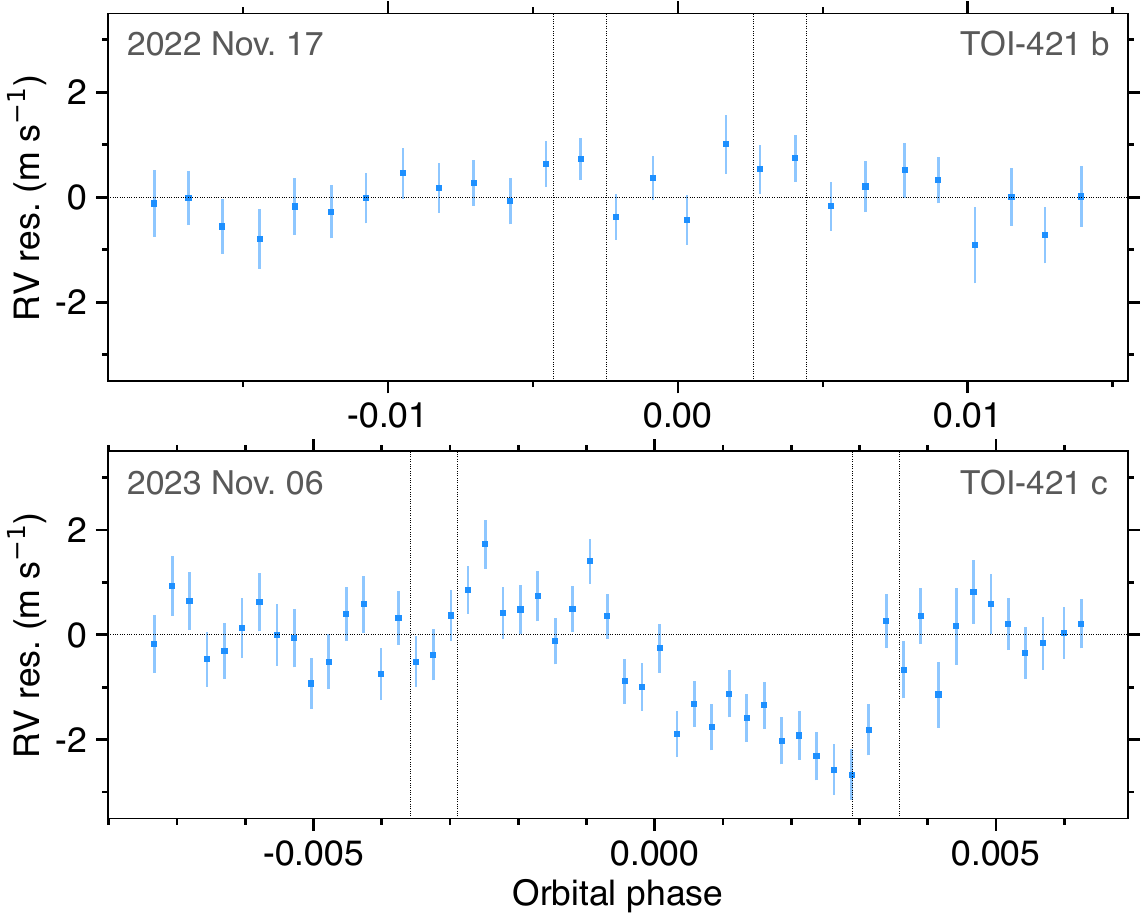}
\centering
\caption[]{RV residuals from the stellar Keplerian motion, phase-folded over the period of TOI-421~b and c. The RM anomaly is visible for the larger planet c. Vertical lines mark transit contacts.} 
\label{fig:RVres}
\end{figure}

ESPRESSO echelle spectra were converted into CCFs with \textsc{antaress} and fitted with Gaussian profiles to assess the quality of their contrast, FWHM, and Keplerian RV residuals. The custom weighted mask built from a master spectrum over the two visits (Sect.~\ref{sec:red_tr_spec}) decreases the dispersion of CCF properties by about 6\% compared to the standard DRS G9 mask, and was adopted hereafter. No systematics were found in the CCF properties in 2022 Nov. 17. The RV residuals and contrast in 2023 Nov. 6 show linear correlations with time and S/N ratio, respectively, which were corrected for in the echelle spectra. Correcting input spectra for sky contamination does not improve the CCF time-series. The RV residuals derived from the detrended datasets, with exquisite uncertainties of $\sim$53\,cm\,s$^{-1}$ on a single measurement, are shown in Fig.~\ref{fig:RVres}. 

\begin{figure}
\includegraphics[trim=0cm 0cm 0cm 0cm,clip=true,width=\columnwidth]{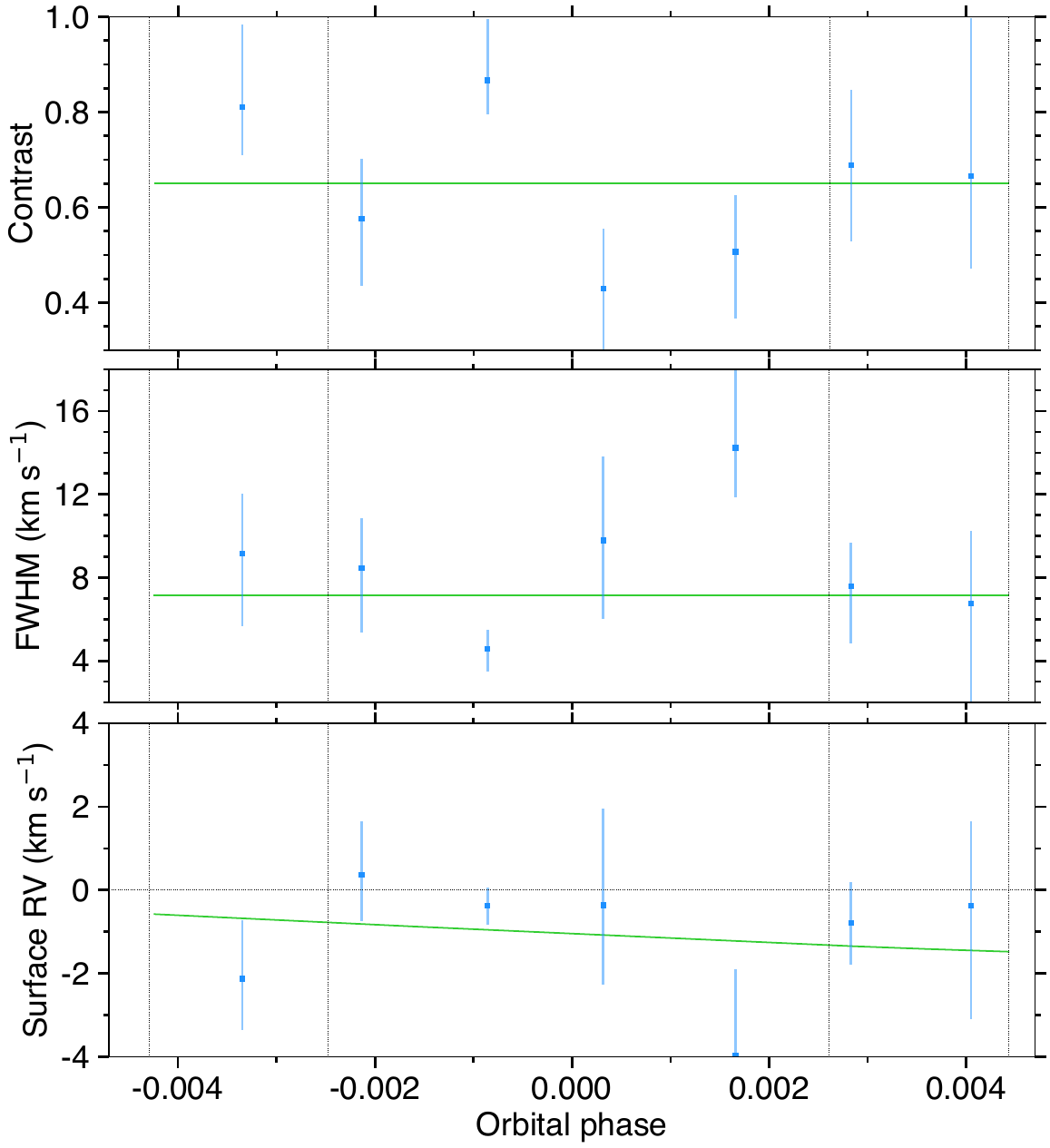}
\centering
\caption[]{Properties of stellar regions occulted by TOI-421~b. Dotted vertical lines show transit contacts. Measured values (blue symbols) are derived from fits to each intrinsic CCF, while green curves are derived from the system joint fit.} 
\label{fig:Intr_prop_20221117}
\end{figure}

\begin{figure}
\includegraphics[trim=0cm 0cm 0cm 0cm,clip=true,width=\columnwidth]{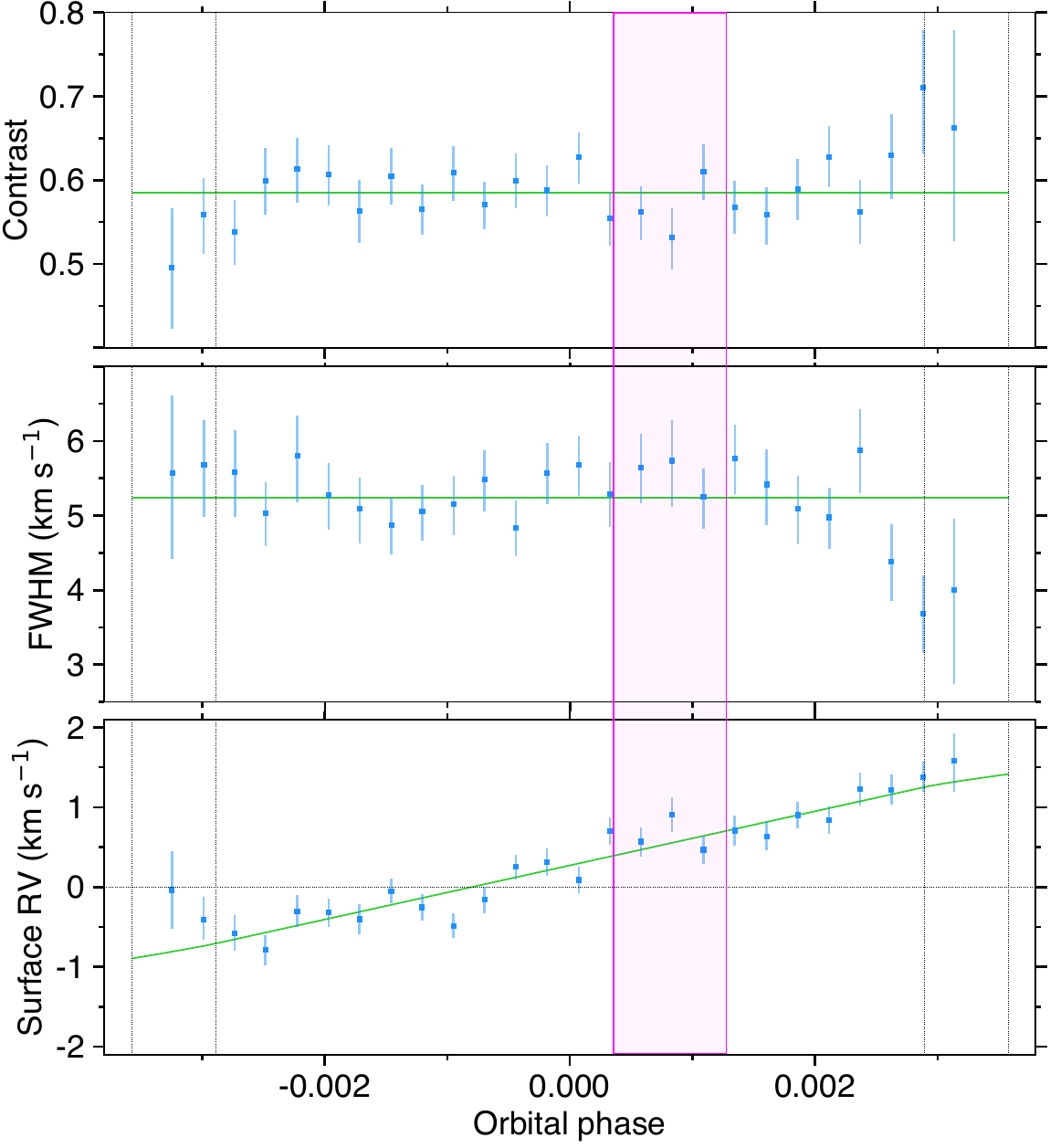}
\centering
\caption[]{Properties of stellar regions occulted by TOI-421~c. Same format as Fig.~\ref{fig:Intr_prop_20221117}. The magenta band hilights the spot crossing window.} 
\label{fig:Intr_prop_20231106}
\end{figure}

Echelle spectra were aligned in the star rest frame, and scaled to their correct relative flux level using a \textsc{batman} lightcurve for TOI-421~b (with the \citealt{Krenn2024} limb-darkening coefficients derived in the CHEOPS passband, which best matches ESPRESSO spectral range), and the \textsc{sage} lightcurve derived from the TOI-421~c photometry. Planet-occulted spectra were then extracted, converted into intrinsic spectra, cross-correlated with TOI-421 custom mask, and analyzed following the RMR procedure (Sect.~\ref{sec:RMR}). Intrinsic line fits to individual exposures were performed using a Gaussian model, with broad uniform priors on the line RV centroid ($\mathcal{U}$(-10,10)\,km\,s$^{-1}$, in case the literature value for $v_{\rm eq} \sin i_\star$ = 1.8\,km\,s$^{-1}$ is underestimated), FWHM ($\mathcal{U}$(0, 18)\,km\,s$^{-1}$, as the local line is narrower than the disk-integrated one with FWHM $\sim$ 6.5\,km\,s$^{-1}$), and contrast ($\mathcal{U}$(0, 1)). The line is not well detected in individual exposures of 2022 Nov. 17 due to the small size of TOI-421~b (Fig.~\ref{fig:Intr_prop_20221117}), so we kept all exposures in the joint fit. The larger size of TOI-421~c allows the line to be clearly detected in all exposures of 2023 Nov. 06, except for the first and last ones that were excluded due to their much lower S/N. The three exposures possibly affected by the spot crossing do not show any clear deviations of their intrinsic properties (Fig.~\ref{fig:Intr_prop_20231106}). Solid-body rotation and a constant line shape across the photosphere are favored in 2023 Nov. 06, and we adopted these models for both visits.

We first performed a joint fit to each visit, with MCMC jump parameters $\lambda$ ($\mathcal{U}$(-180,180)$^{\circ}$), $v_{\rm eq} \sin i_\star$ ($\mathcal{U}$(0,10)\,km\,s$^{-1}$), the contrast ($\mathcal{U}$(0, 1)) and FWHM ($\mathcal{U}$(0, 18)\,km\,s$^{-1}$) of the intrinsic stellar line. 
The fit to the TOI-421~b visit yields a detection for the planet-occulted stellar line, with well-defined PDFs for its FWHM and contrast. However, the transit chord is poorly constrained due to the small transit depth and grazing orbit. The PDF for $v_{\rm eq} \sin i_\star$ peaks at the expected value but is consistent with 0\,km\,s$^{-1}$, and as a result the PDF for $\lambda_\mathrm{b}$ is ill-defined, with a peak at -150$^{\circ}$ and a broad wing that extends up to about 0$^{\circ}$. 
The fit to the TOI-421~c visit was first performed without spotted exposures, yielding clean measurements for all properties. We checked that more complex surface RV models and CLV of the intrinsic line were not justified by the data. We then fixed the derived properties and used a spotted RMR model (Mercier et al., in prep.) to fit the spot latitude, flux contrast, angular aperture, and time of inferior conjunction. 
In contrast to photometry, the spectroscopic data of TOI-421 does not appear to be sensitive to the spot, yielding uniform PDFs for its properties. Assuming the spot and quiet star display the same line shape and RV field, the average line from spotted exposures differs through the fraction of occulted spot area, which is fainter and has a smaller brightness-weighted RV. This distortion is likely negligible for TOI-421~c due to its small surface, low contrast of the spot, and slow stellar rotation. We thus performed subsequent fits with an unspotted model, including spotted exposures. We derive $v_{\rm eq} \sin i_\star$ = 1.60$\pm$0.09\,km\,s$^{-1}$, consistent with the spectroscopic value from \citealt{Carleo2020} (1.8$\pm$1.0\,km\,s$^{-1}$) and $\lambda_\mathrm{c}$ = 14.0$\pm$1.8$^{\circ}$.

\begin{figure}
\includegraphics[trim=0cm 0cm 0cm 0cm,clip=true,width=\columnwidth]{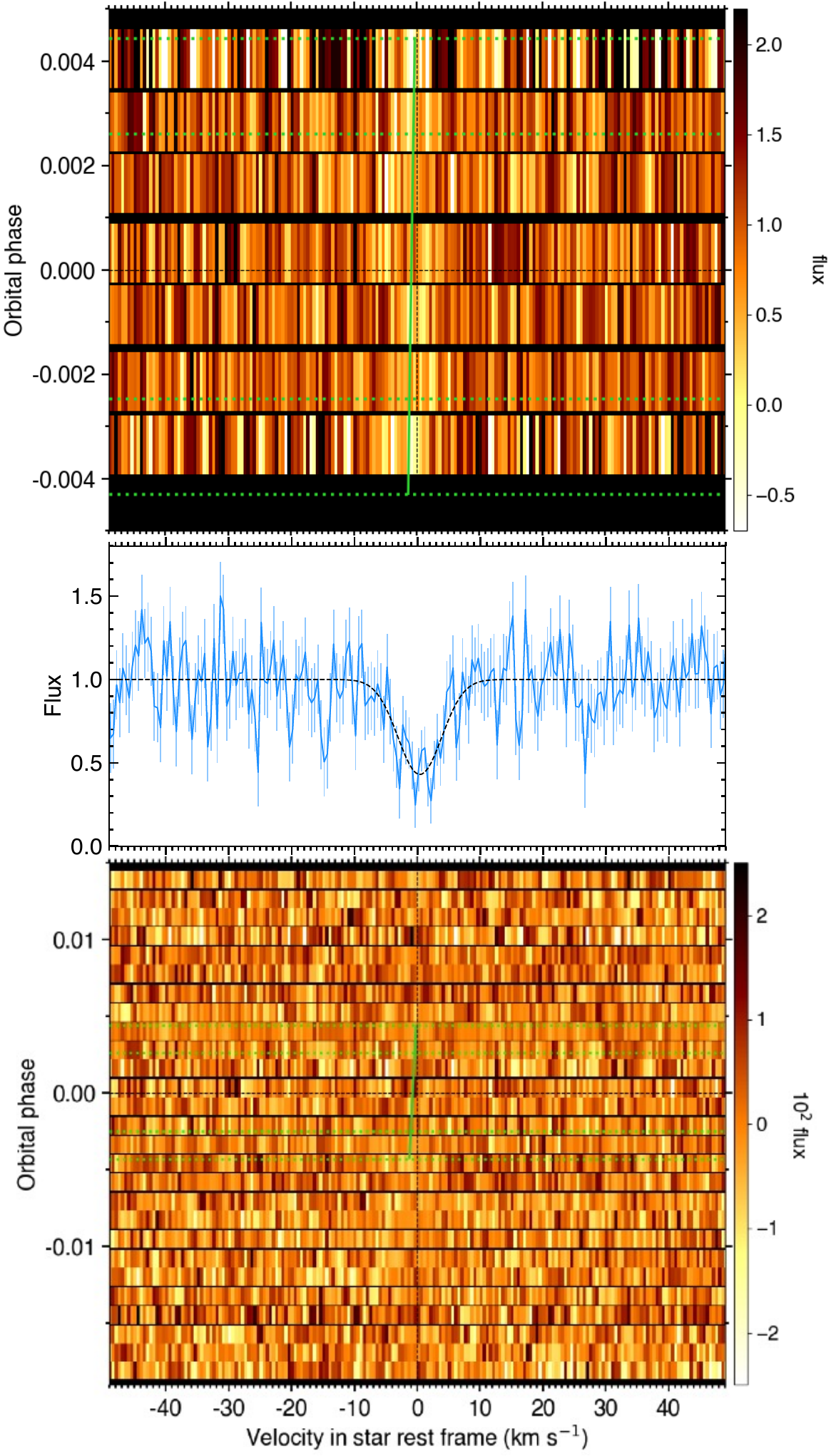}
\centering
\caption[]{Intrinsic CCFs occulted by TOI-421~b. \textit{Top:} Flux map, as a function of RV in the star rest frame (in abscissa) and orbital phase (in ordinate). The core of the stellar line is barely visible in individual exposures, but detectable through the joint RMR fit. The green solid line shows the best-fit surface RV model. Transit contacts are shown as green dashed lines. \textit{Middle:} Weighted average of intrinsic CCFs aligned in a common rest frame, with a Gaussian fit to highlight the line shape. \textit{Bottom:} Residuals between the disk-integrated master-out and individual CCFs (outside of transit) and between intrinsic CCFs and their best-fit RMR model (during transit).} 
\label{fig:Intr_maps_b}
\end{figure}

\begin{figure}
\includegraphics[trim=0cm 0cm 0cm 0cm,clip=true,width=\columnwidth]{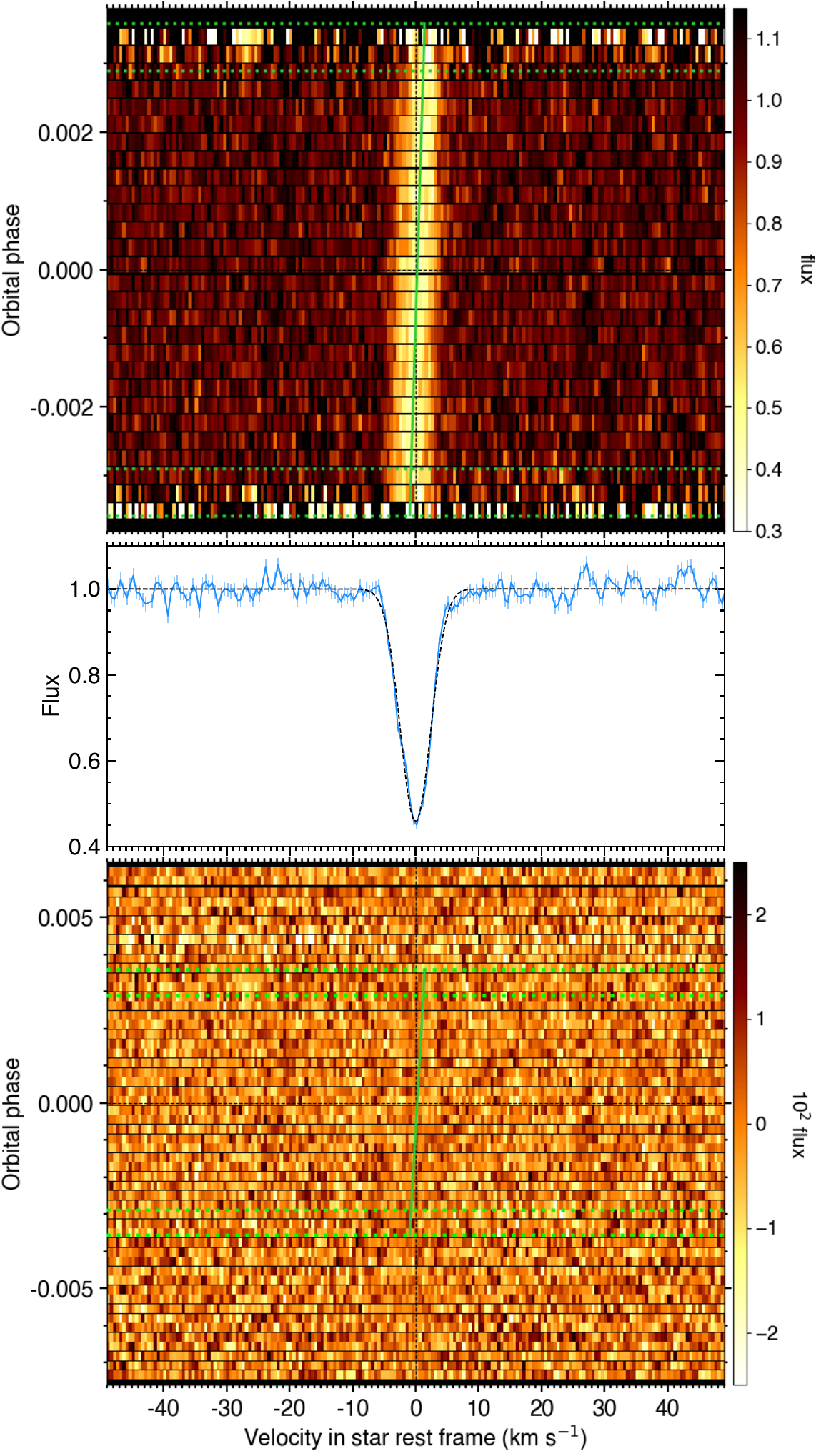}
\centering
\caption[]{Intrinsic CCFs occulted by TOI-421~c. Same format as in Fig.~\ref{fig:Intr_maps_b}. Due to the larger planet size, the occulted line track is clearly visible.} 
\label{fig:Intr_maps_c}
\end{figure}

We then performed a joint fit to both visits, using a common $v_{\rm eq} \sin i_\star$. The fit is of good quality, with a reduced $\chi^2$ unity and no strong systematic features in the residuals (Fig.~\ref{fig:Intr_maps_b} and \ref{fig:Intr_maps_c}). Results for $v_{\rm eq} \sin i_\star$ and $\lambda_\mathrm{c}$ remain the same as for the best fit to the TOI-421~c dataset. On the other hand, the constraint brought by this dataset on $v_{\rm eq} \sin i_\star$ refines the PDF of $\lambda_\mathrm{b}$, so that it now shows two modes at about -150 and -35$^{\circ}$ (Fig.~\ref{fig:Corr_diag_TOI421}). Although the anti-aligned configuration yields the stronger mode we consider it dynamically unlikely and isolated the second mode, which yields $\lambda_\mathrm{b}$ = -35$\stackrel{+26}{_{-23}}$ $^{\circ}$. The mutual inclination between TOI-421~b and c is then $i_\mathrm{bc}$ = 35$\stackrel{+14}{_{-23}}$ $^{\circ}$, showing that their orbital planes are not significantly misaligned but may still not be perfectly coplanar.

Both \citet{Carleo2020} and \citet{Krenn2024} estimated a stellar rotation period of 39.6$\pm$1.6\,days from the modulation measured in RVs, activity indexes, and TESS photometry. Furthermore, widespread activity-rotation relationships (\citealt{Mamajek2008,Suarez2016}) yield a similar period from the mean log $R'_\mathrm{HK}\sim$-4.9. However, this corresponds to a maximum $v_{\rm eq} \sin i_\star$ of about 1.1\,km\,s$^{-1}$, which is not consistent with our RM-derived value (1.6\,km\,s$^{-1}$) nor, in fact, with the spectroscopic value from \citet{Carleo2020} (1.8\,km\,s$^{-1}$). 
Interestingly, the rotation distribution of GKM dwarfs is known to be bimodal (e.g \citealt{Nielsen2013,McQuillan2014,Davenport2018,Santos2021}). The most plausible origin is a momentary stall of the stellar surface spin-down, due to the angular momentum transfer between the fast-rotating core and slow-rotating envelope (\citealt{McQuillan2014,Angus2020,Spada2020,Gordon2021,Lu2022,Santos2024}). Accordingly this rotation gap is not observed for fully convective stars, but is found in partially convective stars and is most prominent for K and M dwarfs. This discontinuity is also present in activity-rotation diagrams (\citealt{Santos2023}), yet it is not accounted for even in recent relations between period and log $R'_\mathrm{HK}$  (\citealt{Suarez2016,Suarez2018}). At log $R'_\mathrm{HK}$ $\sim$-5, the sample of G-K stars upon which these relations are built display a broad range of periods between $\sim$15-50 days. The activity-derived rotation period of TOI-421 (G9-type) 
is thus likely biased, and the apparent discrepancy between the period derived from the observed modulation and the spectroscopic- and RM-derived limits is solved if the former ($\sim$40\,days) traces the second harmonic of the true rotation ($\sim$20\,days).

\begin{figure}
\includegraphics[trim=0cm 0cm 0cm 0cm,clip=true,width=\columnwidth]{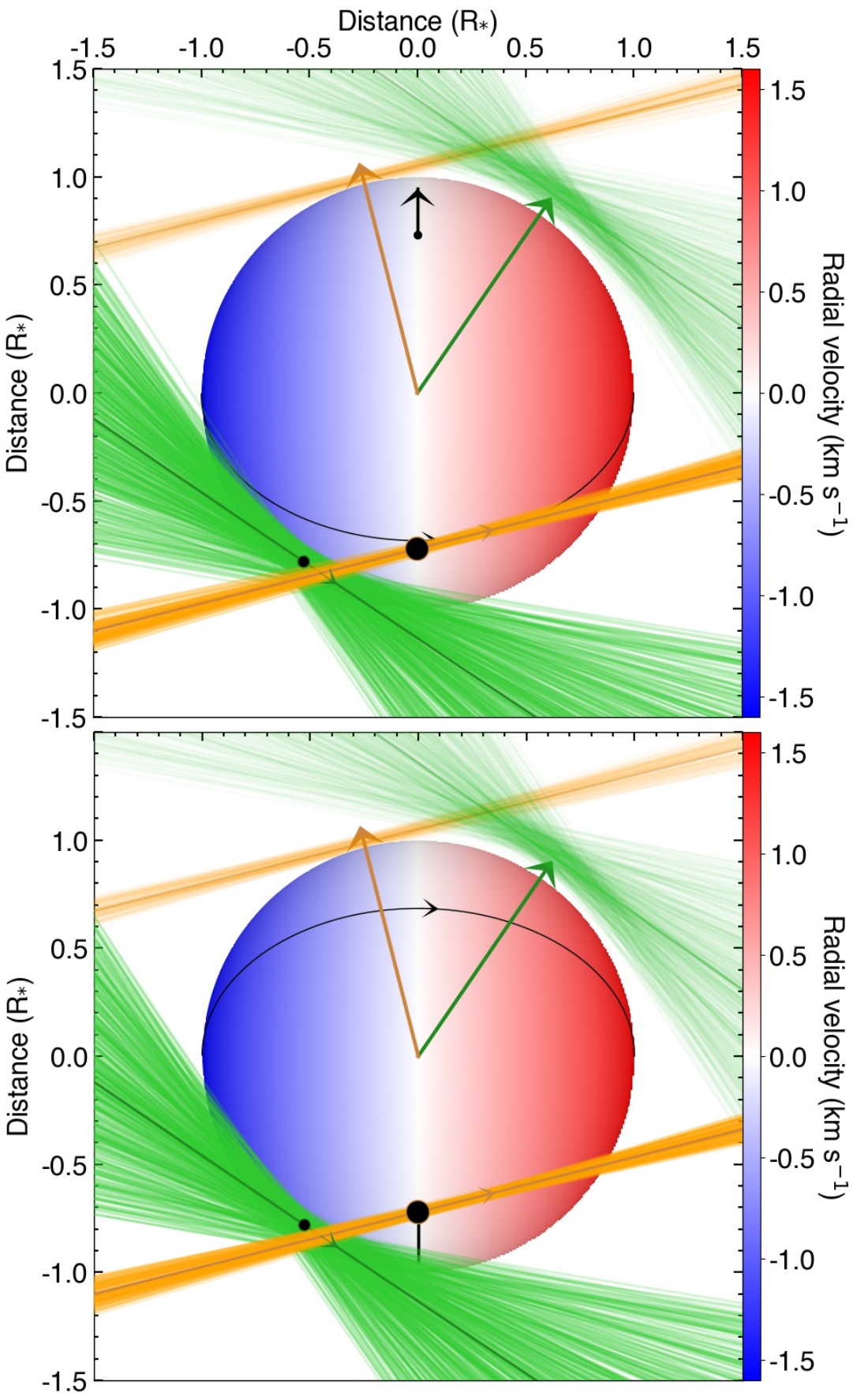}
\centering
\caption[]{Best-fit architecture of the TOI-421 system, for the Northern (top) and Southern (bottom) configurations. The stellar spin--axis is the black arrow extending from the north pole, with the stellar equator plotted as a solid black line. The stellar disk is colored as a function of its surface RV field. The green and orange solid curves represent the best-fit orbital trajectory for TOI-421~b and c, with normals shown as arrows extending from the star center (the planets do not undergo stellar occultation due to the eccentricity of their orbits). The thinner colored lines show orbits obtained for orbital inclination, semi-major axis, and sky-projected spin-orbit angle values drawn randomly within 1$\sigma$ from their PDFs. The star, planets (black disk), and orbits are to scale. We note that the two planets were not simultaneously transiting in our observations.} 
\label{fig:system}
\end{figure}

We thus performed the final joint fit with jump parameters $\cos i_\star$ (with uniform prior $\mathcal{U}$(-1, 1)), $R_\star$ (with normal prior $\mathcal{N}$(0.866,0.006)\,$R_\odot$), and $P_\mathrm{eq}$ ($\mathcal{N}$(19.8,0.8)\,days). The resulting correlation diagram is shown in Fig.~\ref{fig:Corr_diag_TOI421}, with best-fit properties reported in Table~\ref{tab:TOI421_prop}. The Northern ($i_\star$ = 46.9$\stackrel{+4.2}{_{-4.5}}$ $^{\circ}$ ) and symmetrical Southern configurations for the star (Fig.~\ref{fig:system}) yield similar 3D spin-orbit angles for the two planets, with combined values $\Psi_\mathrm{b}$ = 57.2$\stackrel{+11.2}{_{-15.0}}$ $^{\circ}$ and $\Psi_\mathrm{c}$ = 44.9$\stackrel{+4.4}{_{-4.1}}$ $^{\circ}$.

\subsection{History of the system}

The present-day TOI-421 system is dynamically stable (\citealt{Krenn2024}), albeit with an upper stability limit on its planets' mutual inclination that we estimate at about 60$^{\circ}$ (e.g., \citealt{Correia2005,Correia2010}). If confirmed, the large value that we measured ($i_\mathrm{bc}$ = 35$\stackrel{+14}{_{-23}}$ $^{\circ}$), together with the highly eccentric ($e_\mathrm{b}$ = 0.13$\pm$0.05, $e_\mathrm{c}$ = 0.19$\pm$0.04) and misaligned ($\Psi_\mathrm{b}$ = 57.2$\stackrel{+11.2}{_{-15.0}}$ $^{\circ}$, $\Psi_\mathrm{c}$ = 44.9$\stackrel{+4.4}{_{-4.1}}$ $^{\circ}$) orbits of TOI-421b and c, points toward a chaotic dynamical evolution. TOI-421c may have undergone Kozai-Lidov migration, following resonant cycles with the M dwarf companion of TOI-421 (presently at a separation of $\sim$2200\,au,  \citealt{Mugrauer2020,Carleo2020,Behmard2022}). In that scenario, it is however unclear whether the orbit of TOI-421b would have remained stable, and how it would have reached its present high eccentricity and misalignment. Alternatively, we propose that the planets originally underwent DDM, leading to a compact orbital configuration, followed by HEM, leading to the present-day eccentric and misaligned orbits. This second destabilization phase could be the result of planet-planet scattering (e.g., \citealt{Chatterjee2008,Garzon2022}), possibly involving another planet that was ejected from the system. In that scenario, the planets arrived to their present-day orbit soon after formation and their atmosphere started evolving following the dispersal of the protoplanetary disk. We thus used the \textsc{jade} code to simulate the atmospheric evolution of TOI-421~c, without dynamical evolution. We did not simulate the sub-Neptune TOI-421~b because the current version of the code, with no mixed H/He and water atmosphere, does not allow a proper modeling of its internal structure.

\subsubsection{Stellar evolution}
\label{sec:evol_TOI421}

We used the GENEC code (Sect.~\ref{sec:ev_sim_star}) to compute rotating models of TOI-421 based on the stellar properties from \citet{Krenn2024}. Following our previous works on the evolution of exoplanetary systems, the case of a moderate rotator is considered here, which corresponds to a stellar model computed with an initial rotation rate of 5 $\Omega_{\odot}$ \citep[see e.g.][for more details]{Pezzotti2021}. 
To reproduce a fast present-day rotation of 20\,d for TOI-421, we followed the approach used for K dwarfs in the literature (Sec.~\ref{sec:orb_TOI421} and see e.g \citealt{vanSaders2016}) by assuming that the magnetic braking becomes inefficient above a critical value of the Rossby number (taken here equal to 1).

\subsubsection{Planetary evolution}

\begin{figure}
\includegraphics[trim=0cm 0cm 0cm 0cm,clip=true,width=\columnwidth]{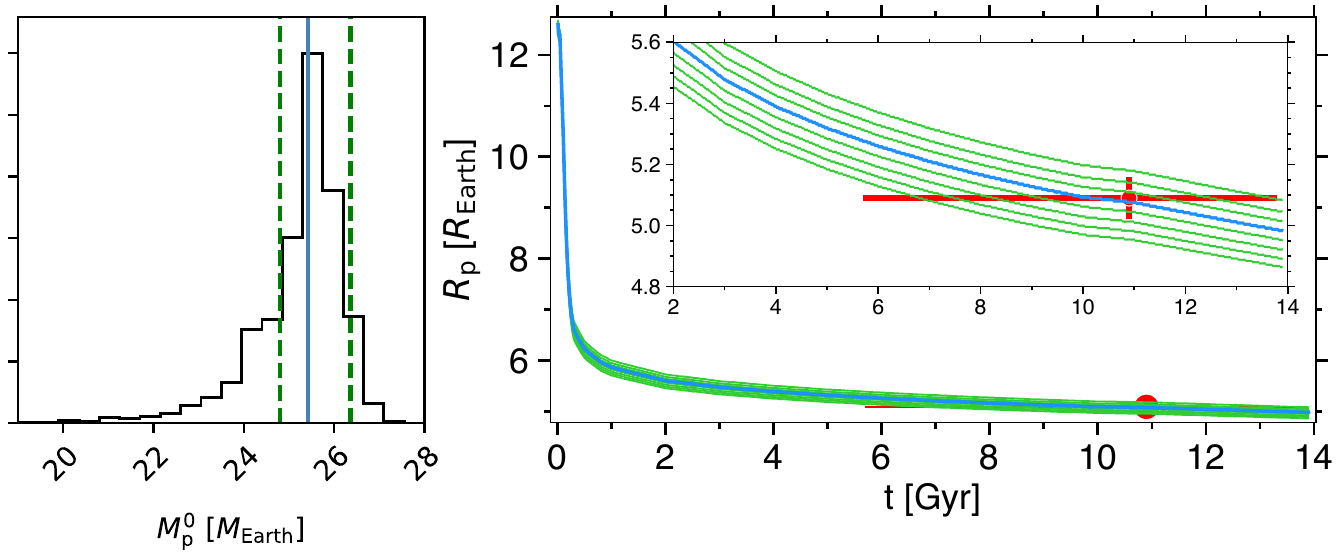}
\centering
\caption[]{\textsc{jade} exploration of TOI-421~c atmospheric evolution. \textit{Left:} PDF of the initial planet mass, with the median highlighted as a blue line and 1$\sigma$ HDI interval as dashed green lines. \textit{Right}: Temporal evolution of the planet radius, in the best-fit simulation (blue curve) and a set of simulations from within the HDI interval (green curves). The red point indicates the measured age and radius of the planet, with their associated uncertainties.}  
\label{fig:Rp_time}
\end{figure}

We simulated the atmospheric evolution of TOI-421~c with \textsc{jade} over a grid of initial planet mass, running the simulations from the expected dissipation of the disk (10\,Myr) up to 14\,Gyr to account for the uncertainty on the system age. System properties were fixed to their present-day values from Table~\ref{tab:TOI421_prop}, and to the results of the internal structure retrieval. The stellar luminosity curve was set to the model computed with \textsc{GENEC}. The evolutionary retrieval was performed as described in Sect.~\ref{sec:ev_sim_pl}, constrained by the present-day planet radius and system age. Results are shown in Fig.~\ref{fig:Rp_time}. At fixed orbital distance the evolution of the planetary atmosphere is dominated by that of the star. Most of the atmosphere is lost within the first hundreds million years when the star is the most energetic, but the inefficient magnetic braking and corresponding high XUV emission of TOI-421 results in persistent mass loss up until present day. In this scenario, we find that TOI-421~c most likely formed with an initial envelope mass $M_\mathrm{Env}^\mathrm{0}$ = 13.3$\stackrel{+0.9}{_{-0.6}}$\,M$_\oplus$, corresponding to a mass fraction of about 0.5 relative to the initial planet mass $M_\mathrm{p}^\mathrm{0}$ = 25.4\,M$_\oplus$. These values are larger than those derived by \citet{Krenn2024} ($f_\mathrm{Env/Pl}^\mathrm{0} \sim$ 0.32--0.42; $M_\mathrm{p}^\mathrm{0} \sim$ 15.5--18 M$_\oplus$) from MESA simulations assuming a fixed stellar evolution with rotation rate of 1\,d at 150\,Myr, comparable to 2\,d for our GENEC model. 
The main difference lies in our later evolution for the star, with luminosities about an order of magnitude larger when assuming a weaker braking that leads to a stronger atmospheric erosion.  
We also note that we neglect the compression of the envelope through self-gravity, which would give it a smaller radius and may become important for mass fractions larger than $\sim$0.5 (\citealt{Owen2017}).
As mentioned by \citealt{Krenn2024} their own models are not ideal to simulate the atmospheric evolution of the TOI-421 system. Our results should similarly be considered as indicative of the trends underlying the planet evolution rather than for their absolute values.

\section{Discussion and perspectives}
\label{sec:discuss}

The exo-Neptunian landscape is a window into formation and evolution processes acting on exoplanets. We proposed a simplified but comprehensive scenario to explain the observed distribution of close-in Neptune properties, in which they form beyond the ice line as two populations distinct in density. Most fluffy Neptunes would be spread by DDM across a broad range of orbital periods, maintaining their primordial orbits, while most dense Neptunes would be brought through HEM to the Ridge and the Desert, acquiring eccentric and misaligned orbits. 

Since the seminal detections of mass loss and a misaligned architecture for the warm Neptune GJ\,436b (\citealt{Ehrenreich2015,Bourrier_2018_Nat}), the existence of hot Neptunes on polar orbits has been connected to the origins of the Desert and to specific dynamical and atmospheric pathways (\citealt{Owen2018,Bourrier2023}). We propose that the Ridge is a hotspot for evolutionary processes, acting as a focus for the HEM of close-in Neptunes, the polarization of their orbits, and their atmospheric erosion. In parallel to our study, \citet{Handley2024} also noted that polar Neptunes are on orbits shorter than 6 days (\textit{i.e.}, within the Ridge) with high planet-to-star mass ratio. This correlation may derive from the one we proposed with planetary density, or may trace processes linked to the formation of Neptunes rather than their migration (see also \citealt{Chen2024}).

Whatever their origin, fluffy Neptunes do not appear resilient enough to survive within the Desert. They would fully erode below a density-period brink within the Ridge, turning into smaller sub-Neptunes or rocky cores that are not part of the exo-Neptunian landscape. Fluffy Neptunes detected at shorter periods than the brink may thus trace late Neptunian migrators, which survived erosion because it was recently triggered by their HEM. The Desert appears to be mainly populated by dense Neptunes, and we propose that they formed and migrated differently from fluffy Neptunes, possibly explaining the correlation observed between Desert planets and the higher metallicity of their host star (\citealt{Vissapragada2025}). This is also backed-up by the work of \cite{Doyle2025b} who completed a statistical analysis on the complete population of Neptunian desert planets and found those within the Desert do not possess high envelope mass fractions compared to planets within the Ridge.


Validating a complete theory of Neptunian origins requires further detection of these planets at all periods, with precise mass and radius to better map the landscape and assess the existence of features like the brink. It also requires gathering large samples of mass loss and orbital architecture measurements, to determine the relative roles of evaporation and the different migration pathways. The consortium of the NIRPS spectrograph (Bouchy et al. 2025) has started to acquire helium measurements in a sample of close-in Neptunes (Allart et al. 2025), which provide constraints on their mass loss when compared with numerical simulations such as EvE. These precise measurements will soon be complemented by the large-scale survey from the NIGHT spectrograph, dedicated to the search for helium (\citealt{FarretJentink2024_proc,FarretJentink2024}). Similarly, ATREIDES aims at carrying out a survey of orbital architectures and optical atmospheric measurements, focusing on the close-in Neptune population.  


We described in detail the observational strategy of ATREIDES to guide future large transit programs. Spectroscopic transit observations benefit from simultaneous transit photometry to better understand stellar contamination and characterize the planetary atmosphere, from contemporaneous transit photometry to revise transit ephemeris, and from long-term photometry to characterize the stellar rotation and activity. We also described the methodology that will be used to exploit the transmission spectroscopy datasets, and that we propose as a standard approach to ensure robust and reproducible results. It combines a careful and detailed reduction of the spectra with the \textsc{ANTARESS} pipeline, and their interpretation with dedicated models such as EvE to disentangle stellar and planetary atmospheric signatures, and analyze the RM effect. The present-day properties of the planetary system can then inform evolutionary models of the star and planet, such as \textsc{genec} and \textsc{jade}, to rewind the system history. As an illustration of this approach we analyzed the first ATREIDES dataset, combined with archival ESPRESSO data, to derive the orbital architectures of the Neptune TOI-421~c and sub-Neptune TOI-421~b. Both planets have eccentric orbits ($\sim$0.15), which we find to be highly misaligned with the star (about 50$^{\circ}$) and possibly far from coplanarity (about 35$^{\circ}$). We propose that the planets were brought close-in by DDM, followed by a destabilizing phase that reshaped the 3D orbital architecture of the system to its present configuration. In that scenario the planets would have started eroding early-on from the outset of the DDM phase. Our measurement of the stellar rotation period suggests that it underwent weak magnetic braking, maintaining a high luminosity level that would have eroded up to about 85\% of TOI-421~c atmosphere and led to its current mass fraction of about 15-20\% relative to the whole planet mass. Follow-up of the M dwarf companion will help better understand its influence of the dynamical origin of the system, especially if its orbit is found to be eccentric and/or misaligned with the primary host. We also estimate that the nodal precession period of TOI-421b orbital plane is on the order of 1600\,yrs, which should lead to changes in transit duration that could be detected with high-precision photometry over several years of monitoring.


We provide a public catalog associated with ATREIDES targets\footnote{Available on the DACE platform at \url{https://dace.unige.ch/exoplanets/?catalog=atreides}}, which already stores their revised ephemeris to enable follow-up observations. The catalog will be updated as systems are progressively studied, storing other key stellar and planetary properties, as well as \textsc{ANTARESS} homogeneous spectroscopic data products. In particular, we will build a library of 1D disk-integrated and local spectra covering the wide range of ATREIDES stellar properties, which can be used to validate and improve stellar atmospheric models, and to characterize center-to-limb variations of specific spectral lines. The atmosphere of Neptune-size planets remain difficult to probe at optical wavelengths, so that the constraints we will set on absorption from critical tracers (sodium, potassium, water) will provide the community with a list of Neptune candidates to follow-up with the JWST and ground-based spectrographs, and inform the feasibility of larger surveys.


ATREIDES is intended to drive a community effort aimed at understanding the dynamical and atmospheric evolution of close-in Neptunes. We welcome new collaborators interested in the homogeneous processing of their ESPRESSO datasets on Neptunes, in complementary interpretations of the studied systems, and in their follow-up with velocimetry, photometry, astrometry, and imaging. The search for massive outer companions will be essential to correlate misalignments and HEM. We encourage new orbital and atmospheric measurements across the exo-Neptunian landscape, in particular within the Ridge. Large samples are necessary to carry out statistical studies of the Neptunian population, and to constrain population synthesis simulating not only for DDM, but also the long-term coupling between HEM and evaporation. 

On the longer-term we plan to link our exploration of the close-in Neptune landscape with that of disk architectures (as can be probed, e.g., with ALMA, \citealt{Czekala2019}), to determine how they compare Neptunes undergoing DDM, assumed to maintain their primordial orbital architectures. This link with formation will also be made through the study of atmospheric composition, using e.g. the JWST, as Neptunes that migrated early-on toward the inner disk are expected to be enriched in heavy elements.


\begin{acknowledgements}
We thank the anonymous reviewer for their valiant reading of this paper and appreciative feedback on our work.

We thank A. Krenn for sharing complementary information about the TOI-421 system.

We thank ESO for supporting ATREIDES and proposing efficient ways to carry out the program. 

This project has received funding from the European Research Council (ERC) under the European Union's Horizon 2020 research and innovation programme (project {\sc Spice Dune}, grant agreement No 947634; project {\sc FIERCE}, grant agreement No 101052347). Views and opinions expressed are however those of the author(s) only and do not necessarily reflect those of the European Union or the European Research Council. Neither the European Union nor the granting authority can be held responsible for them.

SA acknowledges funding from the European Research Council (ERC) under the European Union's Horizon 2020 research and innovation program (Grant agreement No. 865624)

This work has been carried out in the frame of the National Centre for Competence in Research PlanetS supported by the Swiss National Science Foundation (SNSF) under grants 51NF40\_182901 and 51NF40\_205606. 
M.L. acknowledges support of the Swiss National Science Foundation under grant number PCEFP2\_194576. 
K.A. acknowledges support from the Swiss National Science Foundation (SNSF) under the Postdoc Mobility grant P500PT\_230225.
P.E. acknowledges support from the SNF grant No 219745 (Asteroseismology of transport processes for the evolution of stars and planets)
R.A. acknowledges the Swiss National Science Foundation (SNSF) support under the Post-Doc Mobility grant P500PT\_222212 and the support of the Institut Trottier de Recherche sur les Exoplan\`etes (IREx).

A.C.-G. is funded by the Spanish Ministry of Science through MCIN/AEI/10.13039/501100011033 grant PID2019-107061GB-C61. 

W.D. acknowledges the support from FCT - Funda\c{c}\~ao para a Ci\^encia e a Tecnologia through national funds by these grants: UIDB/04434/2020 DOI: 10.54499/UIDB/04434/2020, UIDP/04434/2020 DOI: 10.54499/UIDP/04434/2020.

This research was in part funded by the UKRI Grant EP/X027562/1.
LD would like to acknowledge support from the UKRI (Grant ST/X001121/11).
EG gratefully acknowledges support from UK Research and Innovation (UKRI) under the UK government's Horizon Europe funding guarantee [grant number EP/Z000890/1] and from the UK Science and Technology Facilities Council (STFC; project reference ST/W001047/1).
This work acknowledges funding from a UKRI Future Leader Fellowship (grant numbers MR/S035214/1 and MR/Y011759/1). 

This work is based on data collected under the NGTS project at the ESO La Silla Paranal Observatory. The NGTS facility is operated by a consortium institutes with support from the UK Science and Technology Facilities Council (STFC) under projects ST/M001962/1, ST/S002642/1 and ST/W003163/1.

The contributions at the University of Warwick by BD, SG, PJW have been supported by STFC through consolidated grants ST/T000406/1 and ST/X001121/1.

AP is supported by the \emph{Deut\-sche For\-schungs\-ge\-mein\-schaft, DFG\/} project number PI 2102/1-1

EP acknowledges financial support from the Agencia Estatal de Investigacion of the Ministerio de Ciencia e Innovacion MCIN/AEI/10.13039/501100011033 and the ERDF ``A way of making Europe'' through project PID2021-125627OB-C32, and from the Centre of Excellence ``Severo Ochoa'' award to the Instituto de Astrofisica de Canarias.
S.C.C.B. acknowledges support from FCT with reference 2023.06687.CEECIND/CP2839/CT0002
S.G.S acknowledges the support from FCT through Investigador FCT contract nr. CEECIND/00826/2018 and POPH/FSE (EC).

J.L.-B. is funded by the Spanish Ministry of Science, Innovation and Universities (MCIN/AEI/10.13039/501100011033) through grants PID2019-107061GB-C61, PID2023-150468NB-I00 and CNS2023-144309.

SJM  acknowledges support from the Massachusetts Institute of Technology through the Praecis Presidential Fellowship.

This publication benefits from the support of the French Community of Belgium in the context of the FRIA Doctoral Grant awarded to MT.

F.J.P acknowledges financial support from the Severo Ochoa grant CEX2021-001131-S funded by MCIN/AEI/10.13039/501100011033 and Ministerio de Ciencia e Innovacion through the project PID2022-137241NB-C43.

YGMC acknowledges support from UNAM PAPIIT IG-101224. 

The contributions at the Mullard Space Science Laboratory by E.M.B. have been supported by STFC through the consolidated grant ST/W001136/1.

The postdoctoral fellowship of KB is funded by F.R.S.-FNRS grant T.0109.20 and by the Francqui Foundation.

The ULiege's contribution to SPECULOOS has received funding from the European Research Council under the European Union's Seventh Framework Programme (FP/2007-2013) (grant Agreement n$^\circ$ 336480/SPECULOOS), from the Balzan Prize and Francqui Foundations, from the Belgian Scientific Research Foundation (F.R.S.-FNRS; grant n$^\circ$ T.0109.20), from the University of Liege, and from the ARC grant for Concerted Research Actions financed by the Wallonia-Brussels Federation. 

This work is supported by a grant from the Simons Foundation (PI Queloz, grant number 327127).

J.d.W. and MIT gratefully acknowledge financial support from the Heising-Simons Foundation, Dr. and Mrs. Colin Masson and Dr. Peter A. Gilman for Artemis, the first telescope of the SPECULOOS network situated in Tenerife, Spain.

This work has received fund from the European Research Council (ERC) under the European Union's Horizon 2020 research and innovation programme (grant agreement n$^\circ$ 803193/BEBOP), from the MERAC foundation, and from the Science and Technology Facilities Council (STFC; grant n$^\circ$ ST/S00193X/1). 

TRAPPIST is funded by the Belgian Fund for Scientific Research (Fond National de la Recherche Scientifique, FNRS) under the grant PDR T.0120.21, with the participation of the Swiss National Science Fundation (SNF). 

MG and EJ are F.R.S.-FNRS Research Directors.

XL would like to acknowledge support from the Natural Science Foundation Youth Program of Sichuan Province (grant No. 2024NSFSC1363) and the Doctoral Initiation Fund of West China Normal University (grant No. 22kE036).

\end{acknowledgements}

\bibliographystyle{aa} 
\bibliography{biblio} 

\begin{appendix}

\section{Possible origins of close-in Neptunes}
\label{apn:nept_orig}

\begin{figure*}
\begin{minipage}[tbh!]{\textwidth}
\includegraphics[trim=0cm 0cm 0cm 0cm,clip=true,width=\textwidth]{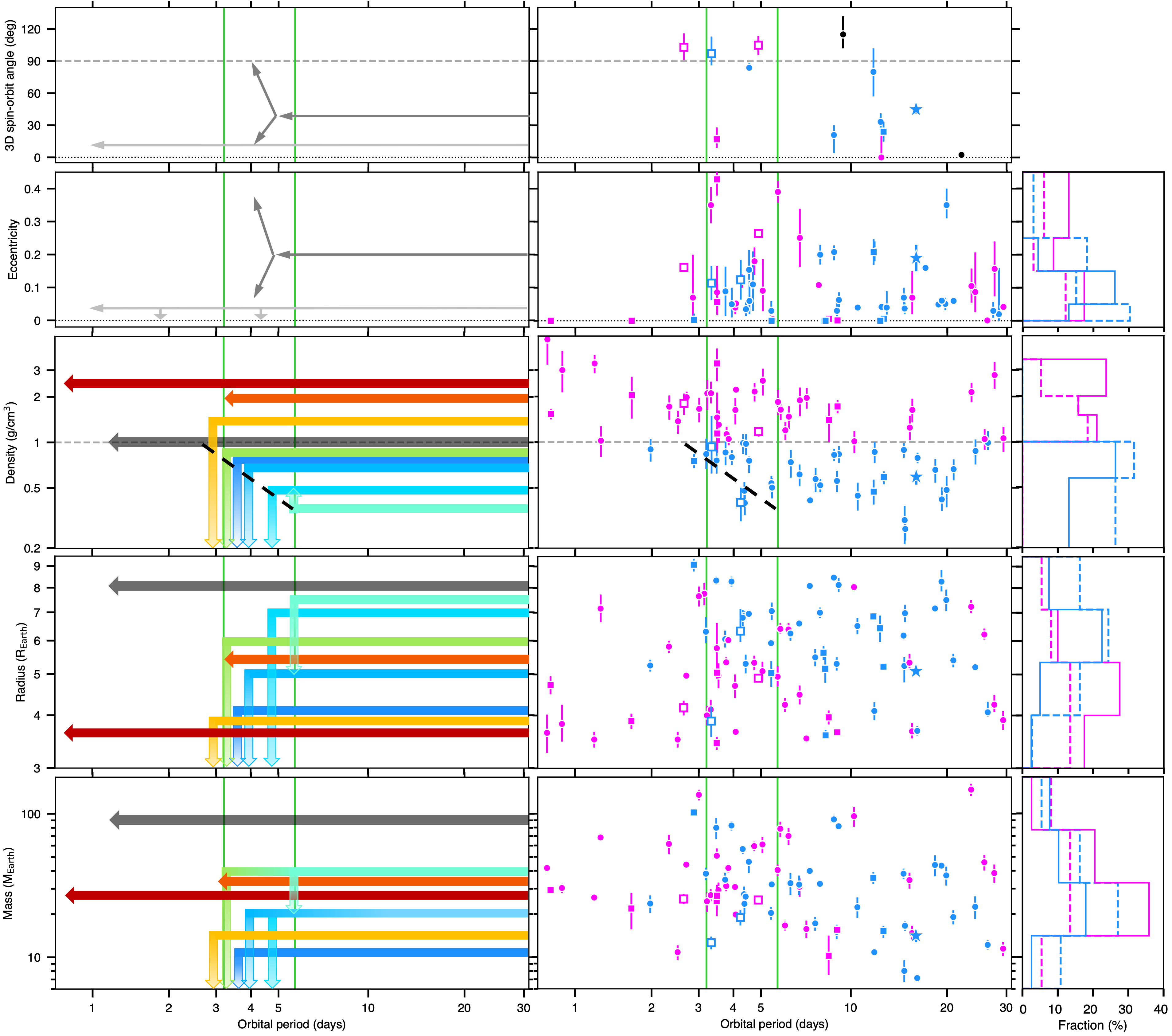}
\centering
\end{minipage}
\caption[]{Close-in Neptune properties as a function of orbital period. Vertical green lines mark the Ridge boundaries. \textit{Middle panels: } Measured properties, colored in blue (resp. magenta) if less dense (resp. denser) than 1\,g\,cm$^{-3}$. 3D spin-orbit angles from \textit{Tepcat} are shown when known to better than 30$^{\circ}$ precision; eccentricity, mass, and radius from the NASA Exoplanet Archive when known to better than 0.1 and 20\%. Density is recomputed from these mass and radius. Star and square symbols have the same meaning as in Fig.~\ref{fig:Per_Rad_PAPER}. \textit{Left panels: } Schematic representation of the proposed atmospheric (broad colored arrows, see text) and dynamical (grey arrows) pathways, as described in the text. The black dashed line in the density diagram marks the approximate location of the proposed density brink. \textit{Right panels:} Except for spin-orbit angles, which have too few measurements, these histograms show the fractions of fluffy and dense planets relative to the Savanna (dashed) or Desert+Ridge (solid) populations. } 
\label{fig:prop_nept}
\end{figure*}

\begin{figure}
\includegraphics[trim=4.5cm 4.5cm 5cm 4.5cm,clip=true,width=0.8\columnwidth]{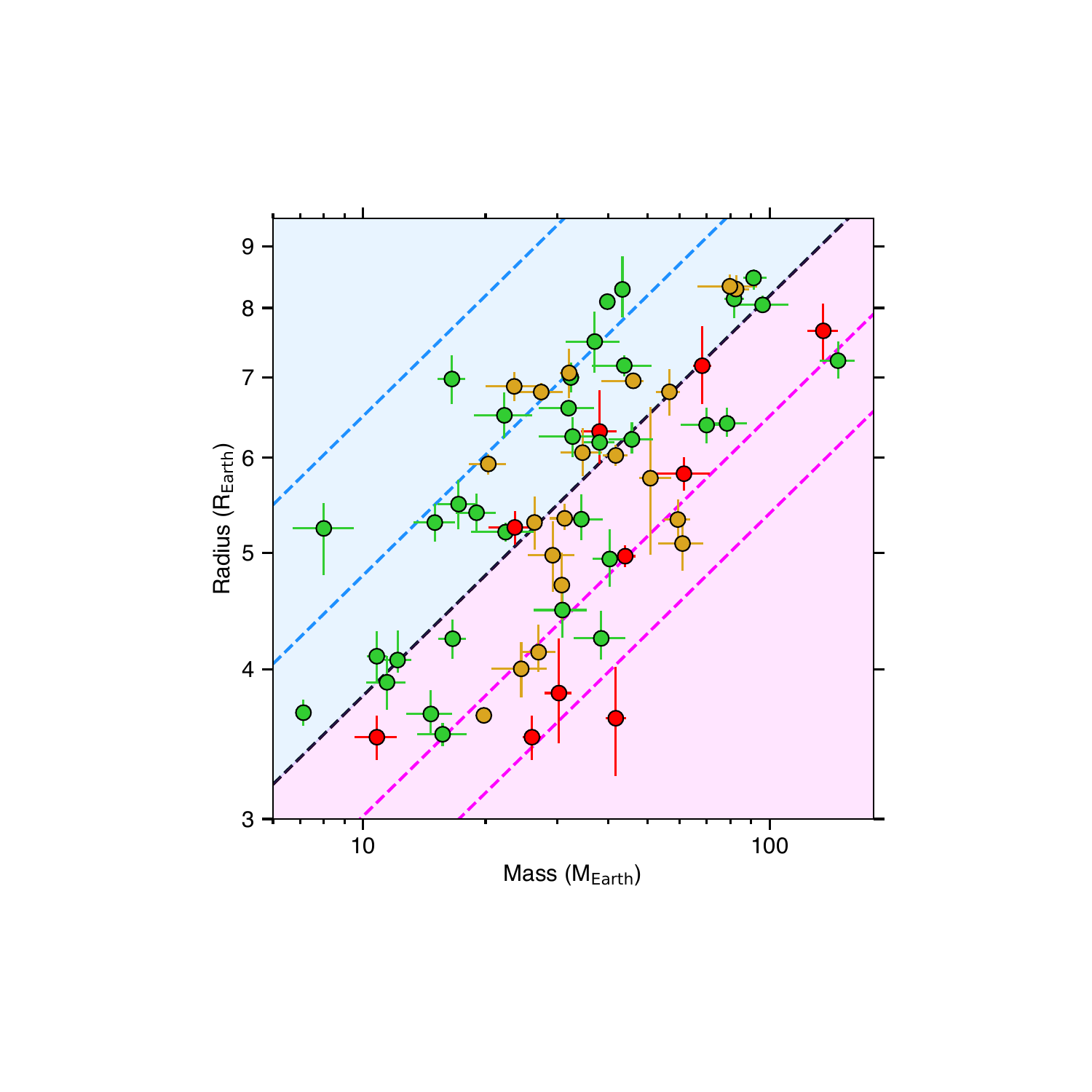}
\centering
\caption[]{Mass-radius diagram of Neptunian planets with mass and radius known to better than with 20\% precision. Green, orange, and red disks corresponds to planets in the Savanna, Ridge, and Desert, respectively. Blue (resp. magenta) background color defines the low-(resp. high) density classes. From left to right, dashed lines indicate iso-density curves at 0.2, 0.5, 1.0, 2.0, 3.5\, g\,cm$^{-3}$} 
\label{fig:MR_nept}
\end{figure}

To draw a consistent picture of the exo-Neptunian landscape we analyzed together the measured distribution of their most relevant properties as a function of orbital period (Fig.~\ref{fig:prop_nept}), together with their radius-mass diagram (Fig.~\ref{fig:MR_nept}). We interpreted these measurements, in particular, using the variation of erosion timescale as a function of atmospheric mass fraction $f_\mathrm{atm}$ proposed by \citet{Owen2018} (heareafter O18), that we summarize below. Neptunian exoplanets are expected to have core masses between $\sim$10-20\,M$_{\oplus}$ (\citealt{Pollack1996,Otegi2020}) and $f_\mathrm{atm}$ larger than 1\%. Above this value, which corresponds to a first stability region (O18), a small increase in the mass of the atmosphere quickly increases its radius and collecting area for stellar irradiation, decreasing the evaporation timescale down to a minimum when the atmosphere mass becomes comparable to the core mass. At this stage, compression of the atmosphere through self-gravity limits its increase in radius with larger mass. Evaporation timescale increases again, leading to the second stability region.

\citet{CastroGonzalez2024b} recently identified a dichotomy between low-density Neptunes within the Savanna ($\rm \sim 0.5 \, g\,cm^{-3}$) and high-density Neptunes within the Ridge ($\rm 1.5-2.0 \, g\,cm^{-3}$). We find in Fig.~\ref{fig:MR_nept} that the lower envelope of the density distribution in the fluffy Neptune regime increases with decreasing orbital periods from the outer Ridge to the inner Desert. We propose that this density ``brink'' traces atmospheric erosion, with a threshold increasing with stronger irradiation. Escaping atmospheres were searched in Neptunes at all orbital periods but were only detected in four of them located within the Ridge (Fig.~\ref{fig:prop_nept}). The non-detection of escape from Neptunes within the Savanna, combined with the absence of clear trend between period and density, suggests that their atmosphere is not strongly hydrodynamically escaping. Thus, we propose that the Ridge marks the threshold for full atmospheric erosion of Neptunian planets. 

We first note the presence of a few Neptunes with both a high radius ($\sim$8\,R$_{\oplus}$) and high mass ($\sim$90\,M$_{\oplus}$), corresponding to $\rho \sim$1\,g\,cm$^{-3}$. They can be found at all periods, from the Savanna into the Desert, suggesting they are resilient to erosion (black arrow in Fig.~\ref{fig:prop_nept}) and can be classified as ``stable'' Neptunes representing a transition between fluffy and dense Neptunes. Then, we note from Fig.~\ref{fig:prop_nept} that there are few Neptunes with M $\approxsup$50\,M$_{\oplus}$ in the Savanna, while more massive Neptunes from both classes are seen in the Ridge. This is a first indication that the Neptunian landscape is shaped by different dynamical processes, with the Ridge as a favored destination for migration.

Fig.~\ref{fig:prop_nept} shows that fluffy Neptunes represent the bulk of the Savanna, displaying a wide range of radii from $\sim$5-7\,R$_{\oplus}$ at $\sim$20\,M$_{\oplus}$ to $\sim$6-8\,R$_{\oplus}$ at $\sim$40\,M$_{\oplus}$ ($\rho \sim$ 0.4-1\,g\,cm$^{-3}$). The Savanna also hosts a few fluffy Neptunes at about 10\,M$_{\oplus}$ and 4\,R$_{\oplus}$ ($\rho \sim$1\,g\,cm$^{-3}$). 
With core masses in between 10-20\,M$_{\oplus}$, Neptunes with mass $\approxinf$20\,M$_{\oplus}$ have $f_\mathrm{atm}<$1, so that larger radii and lower densities correspond to larger $f_\mathrm{atm}$ and faster erosion (O18; blue arrows in Fig.~\ref{fig:prop_nept}). These low-mass Neptunes are absent from the Desert and likely start eroding in the Ridge, turning into stable sub-Neptunes with $f_\mathrm{atm}<$0.01 (O18) or naked cores. 
Neptunes up to $\approxinf$40\,M$_{\oplus}$ have larger $f_\mathrm{atm}$ likely spread around 1. Those with massive core ($\sim$20\,M$_{\oplus}$) are still in the regime where increasing atmospheric mass leads to increasing radius ($f_\mathrm{atm}<$1). They may correspond to the largest and lowest-density 40\,M$_{\oplus}$-Neptunes, which would erode faster than their $\approxinf$20\,M$_{\oplus}$ counterparts and join their evolution track (cyan arrow in Fig.~\ref{fig:prop_nept}). Those with light core ($\sim$10\,M$_{\oplus}$) and $f_\mathrm{atm}>$1 are in the regime where the compressed atmosphere is more stable. They may correspond to the smaller and higher-density 40\,M$_{\oplus}$-Neptunes, which would erode at shorter periods and turn into sub-Neptunes or naked cores (light green arrow in Fig.~\ref{fig:prop_nept}).
These pathways would explain well the proposed density brink, increasing with decreasing period up to $\sim$1\,g\,cm$^{-3}$ (Fig.~\ref{fig:prop_nept}). This scenario requires fluffy Neptunes to be spread across the whole exo-Neptunian landscape at an early age, so that Neptunes migrating left of the density brink would be fully eroded by their energetic young star. This suggests that DDM may be the dominant migration pathway for fluffy Neptunes.

Fig.~\ref{fig:prop_nept} further shows that dense Neptunes are more numerous in the Ridge and Desert than in the Savanna. This is another indication that a fraction of fluffy and dense Neptunes undergo different dynamical evolutions. In the Savanna, the few dense Neptunes appear separated between the massive stable planets, a group with ($\sim$10-20\,M$_{\oplus}$ ; $\sim$3.5-4.5\,R$_{\oplus}$ ; $\rho \sim$1-2\,g\,cm$^{-3}$), and a group with ($\sim$30-40\,M$_{\oplus}$ ; $\sim$4-6\,R$_{\oplus}$ ; $\rho \sim$1-3\,g\,cm$^{-3}$). 
The first group has similar masses as fluffy Neptunes with shallow envelopes but shows smaller radii, and should thus erode at shorter periods due to their lower $f_\mathrm{atm}$ (orange arrow in Fig.~\ref{fig:prop_nept}). Accordingly, there are no dense Neptunes in this mass range within the Ridge and the Desert. These Neptunes likely underwent DDM and major atmospheric erosion like their fluffy counterparts, and may rather be considered as belonging to the same class.
The second group are also smaller and denser than fluffy Neptunes in the same mass range, but since they are likely in the regime of self-compressed atmosphere this corresponds to a larger $f_\mathrm{atm}$. They should thus be closer to the second stability region and more resilient to erosion. Accordingly, dense Neptunes at ($\sim$30-40\,M$_{\oplus}$ ; $\sim$4-6\,R$_{\oplus}$ ; $\rho \sim$1-3\,g\,cm$^{-3}$) are observed in the Ridge, where they are much more numerous than in the Savanna. As noted previously, the Ridge even holds dense Neptunes up to 60\,M$_{\oplus}$ that are not present in the Savanna. It is unlikely that these dense Neptunes in the Ridge are birthed by Neptunes migrating from the Savanna, as inflation would increase their radius and decrease their density, and evaporation would decrease both their mass and radius. Furthermore, the Desert contains dense Neptunes in the $\sim$30-40\,M$_{\oplus}$ range with overall smaller radius ($\approxinf$4Re) and larger density ($\rho\approxsup$2) than in the Ridge, which requires larger cores and less massive envelopes. Again, it is unlikely that these Desert planets result from the erosion of the Ridge ones, as such a substantial decrease in radius would be associated with a large decrease in mass in the self-compressed atmospheric regime. Finally, the mass and radius of the Desert planets do not appear to decrease at short periods, as would be expected from evaporation. Thus, we propose that dense Neptunes with intermediate mass are not shaped by DDM and atmospheric erosion but by HEM, which brings the largest ones preferentially into the Ridge and the smallest ones preferentially into the Desert, presumably because the difference in their core and envelope masses control the dynamical interactions with companions in the system, or trace different formation pathways for these Neptunes.

The eccentricity and 3D spin-orbit angle distributions in Fig.~\ref{fig:prop_nept} yield complementary insights on dynamical evolution. If most Neptunes migrated early-on through DDM, the Ridge should display lower eccentricities, since tidal circularization is more efficient at shorter periods. Yet the proportion of Neptunes on eccentric (0.05 $<$ e $<$ 0.15) orbits increases from the Savanna toward the Ridge, as noted by \citealt{Correia2020}. Furthermore, the Ridge hosts a few Neptunes on highly-eccentric (e$\sim$0.3--0.4) and polar (90$^{\circ}$) orbits, in line with detailed studies that showed a substantial fraction of highly misaligned orbits at the edge of the Desert (\citealt{Albrecht2021,Attia2023,Bourrier2023}). These features, which concern both classes of Neptunes, support the role of HEM in their evolution and further suggest that the Ridge may be preferentially populated by this process. However, HEM would bring Neptunes to highly misaligned orbits, which are expected to be slowly realigned by tidal interactions (\citealt{Fabrycky2007}). Similarly, while HEM would bring Neptunes to eccentric orbits, they should be quickly circularized on the close orbits of the Ridge. The late migration of Neptunes associated with HEM might partly explain residual high eccentricities and misalignment. However, there must be an additional process, such as thermal tides, that counteracts bodily tides (\citealt{Correia2020}) and we propose that another dynamical process would ``polarize'' to 90$^{\circ}$ some of the misaligned orbits. These two processes appear to be specific to the orbital range or Neptunian properties associated to the Ridge.
Furthermore, the Savanna hosts a few Neptunes on mildly-eccentric (e $\sim$0.2) and mildly-misaligned ($\Psi$ $\sim$40$^{\circ}$) orbits, although measurements remain few. A possible scenario could be that HEM brings a small fraction of Neptunes in the Savanna, and a larger fraction into the Ridge, with the former displaying the eccentricity and misalignments acquired at the outcome of HEM, and the latter the orbits further shaped by the aforementioned processes (dark gray arrows in Fig.~\ref{fig:prop_nept}). Meanwhile, DDM would bring most of fluffy Neptunes and a small fraction of dense Neptunes across all orbits, maintaining small eccentricities and misalignments, with only the densest planets surviving early-erosion in the Desert and the Ridge (light gray arrows in Fig.~\ref{fig:prop_nept}).
Interestingly, Jupiter-size planets tend to pile-up in the orbital periods of the Ridge \citep[e.g.][]{Udry2003,Udry2007}, although they display less eccentric orbits than their Neptunian counterparts (\citealt{Correia2020}). A fraction of warm Jupiters in the orbital periods of the Savanna are also on non-circular orbits \citep[e.g.][]{Correia2011}, although this region is less populated than in the Neptunian regime \citep{CastroGonzalez2024a} and shows larger eccentricities. This suggests similar dynamical pathways for Jupiter- and Neptune-size planets \citep[e.g.][]{Dawson2018,Fortney2021}, although they appear to bring Jupiters closer-in and on more eccentric orbits.

\section{Correlation diagrams for the internal structure retrievals}
\label{apn:corr_instr}

\begin{figure*}
\begin{minipage}[h!]{\textwidth}
\includegraphics[trim=0cm 0cm 0cm 0cm,clip=true,width=\columnwidth]{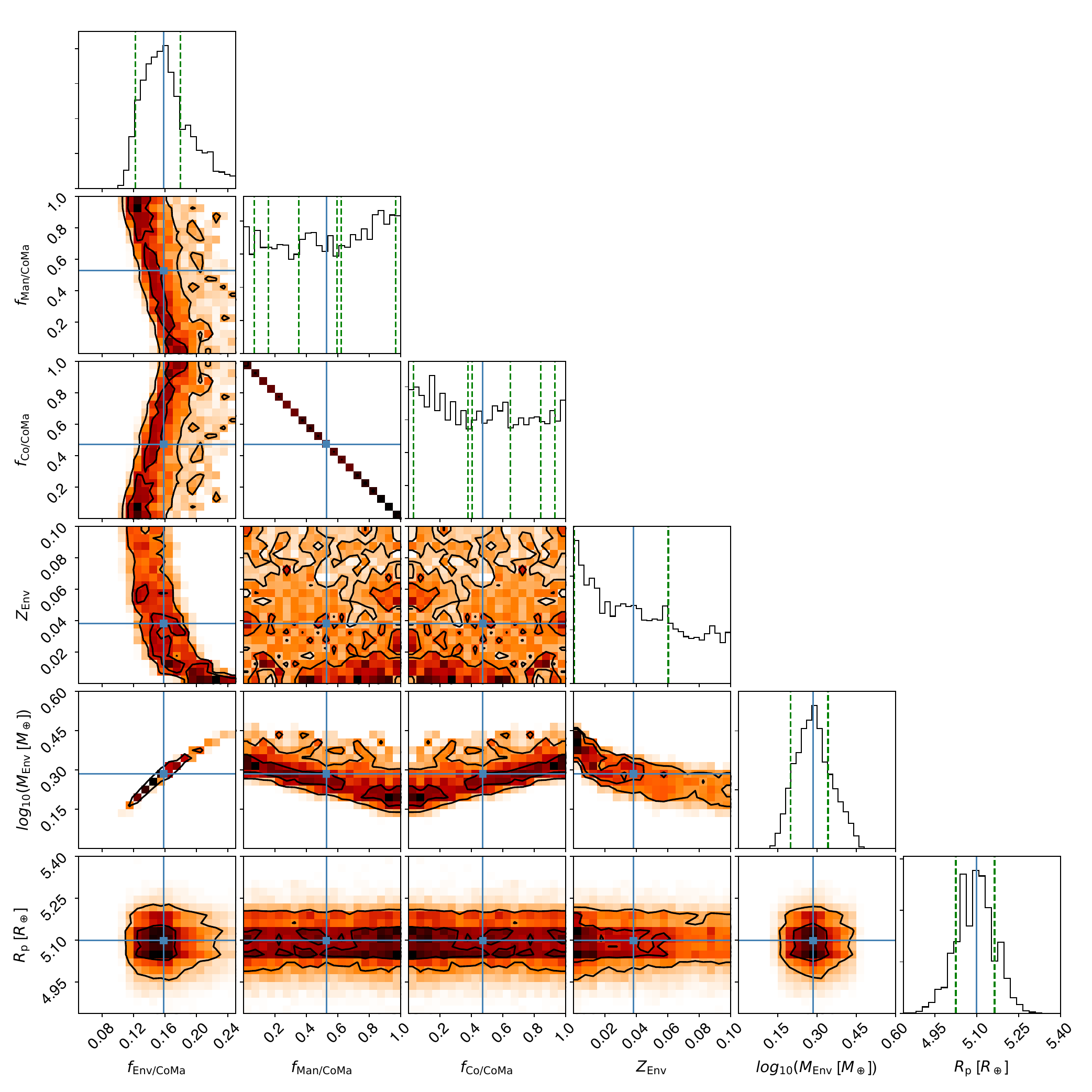}
\centering
\end{minipage}
\caption[]{Correlation diagram for the PDFs of the \textsc{jade} parameters for the interior model of TOI-421~c. The envelope mass, core mass fraction, and planet radius are derived from the other (fitted) parameters. The figure format is the same as in Fig.~\ref{fig:Corr_diag_TOI421}.}
\label{fig:Corr_diag_JADE_int_PAPER}
\end{figure*}

\begin{figure*}
\begin{minipage}[h!]{\textwidth}
\includegraphics[trim=0cm 0cm 0cm 0cm,clip=true,width=\columnwidth]{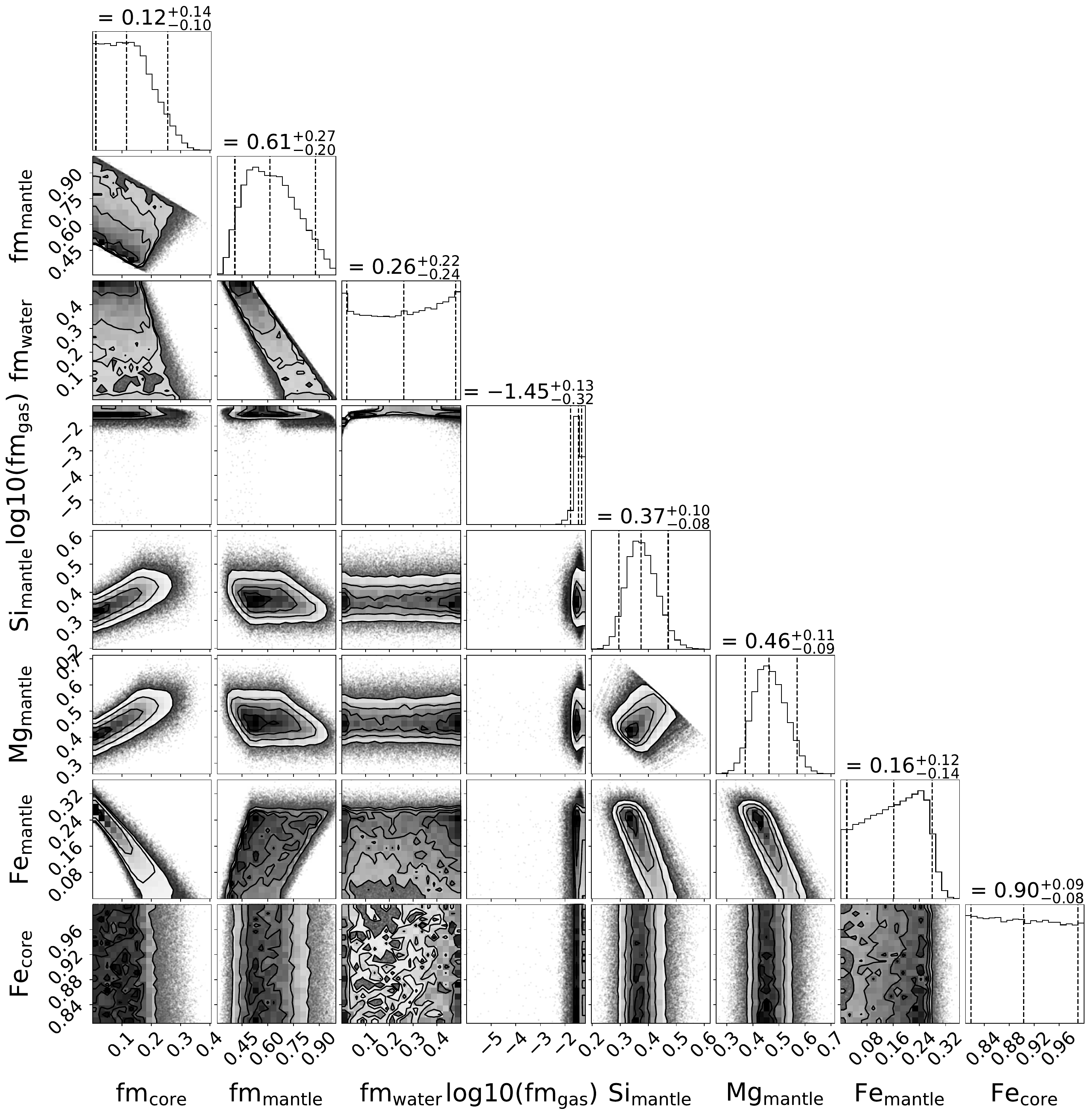}
\centering
\end{minipage}
\caption[]{Correlation diagram for the PDFs of the \textsc{biceps} parameters for the interior model of TOI-421~b (wet case).}
\label{fig:BICEPS_int_b_wet}
\end{figure*}

\begin{figure*}
\begin{minipage}[h!]{\textwidth}
\includegraphics[trim=0cm 0cm 0cm 0cm,clip=true,width=\columnwidth]{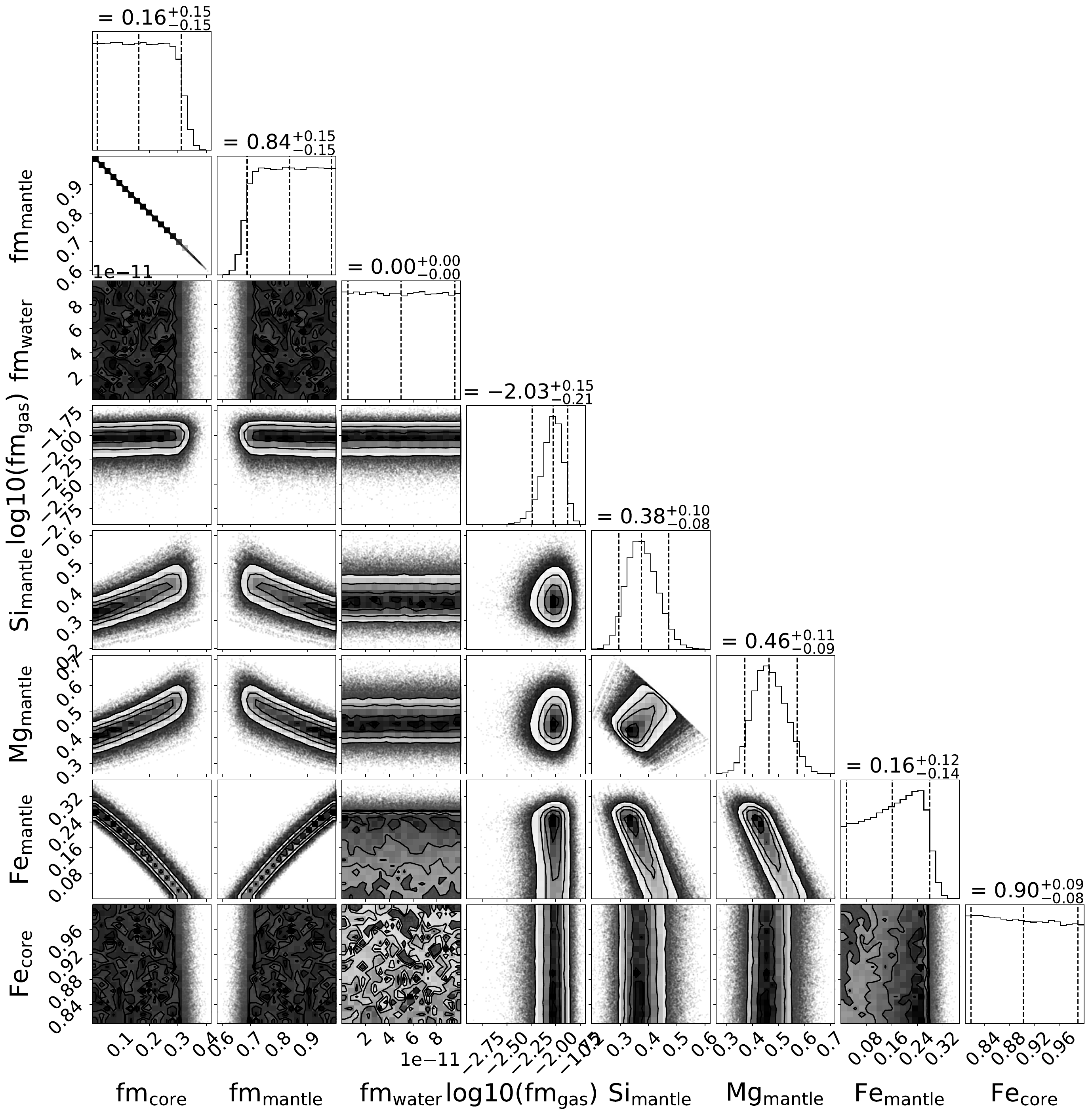}
\centering
\end{minipage}
\caption[]{Correlation diagram for the PDFs of the \textsc{biceps} parameters for the interior model of TOI-421~b (dry case).}
\label{fig:BICEPS_int_b_dry}
\end{figure*}

\begin{figure*}
\begin{minipage}[h!]{\textwidth}
\includegraphics[trim=0cm 0cm 0cm 0cm,clip=true,width=\columnwidth]{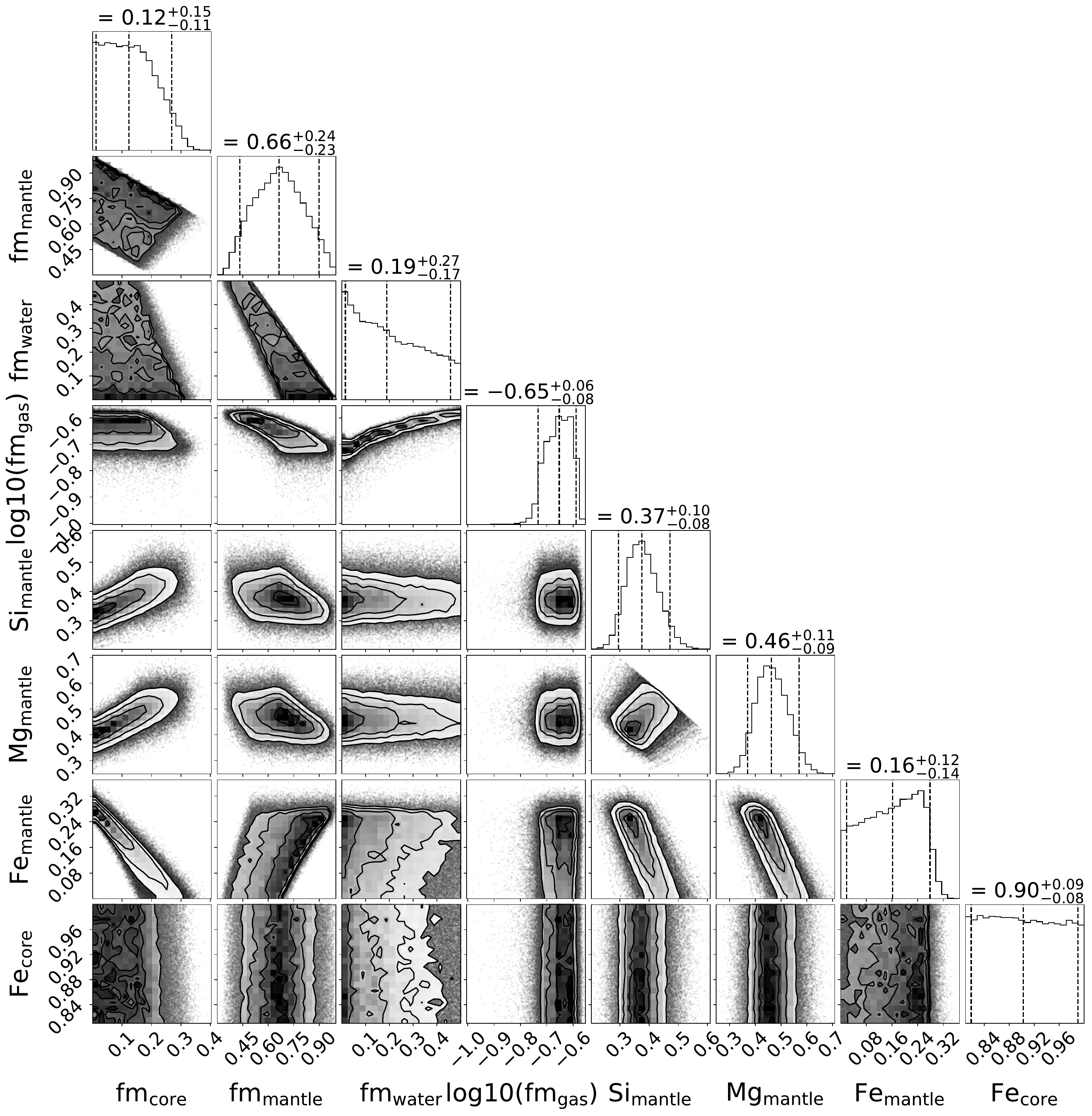}
\centering
\end{minipage}
\caption[]{Correlation diagram for the PDFs of the \textsc{biceps} parameters for the interior model of TOI-421~b (wet case).}
\label{fig:BICEPS_int_c_wet}
\end{figure*}

\begin{figure*}
\begin{minipage}[h!]{\textwidth}
\includegraphics[trim=0cm 0cm 0cm 0cm,clip=true,width=\columnwidth]{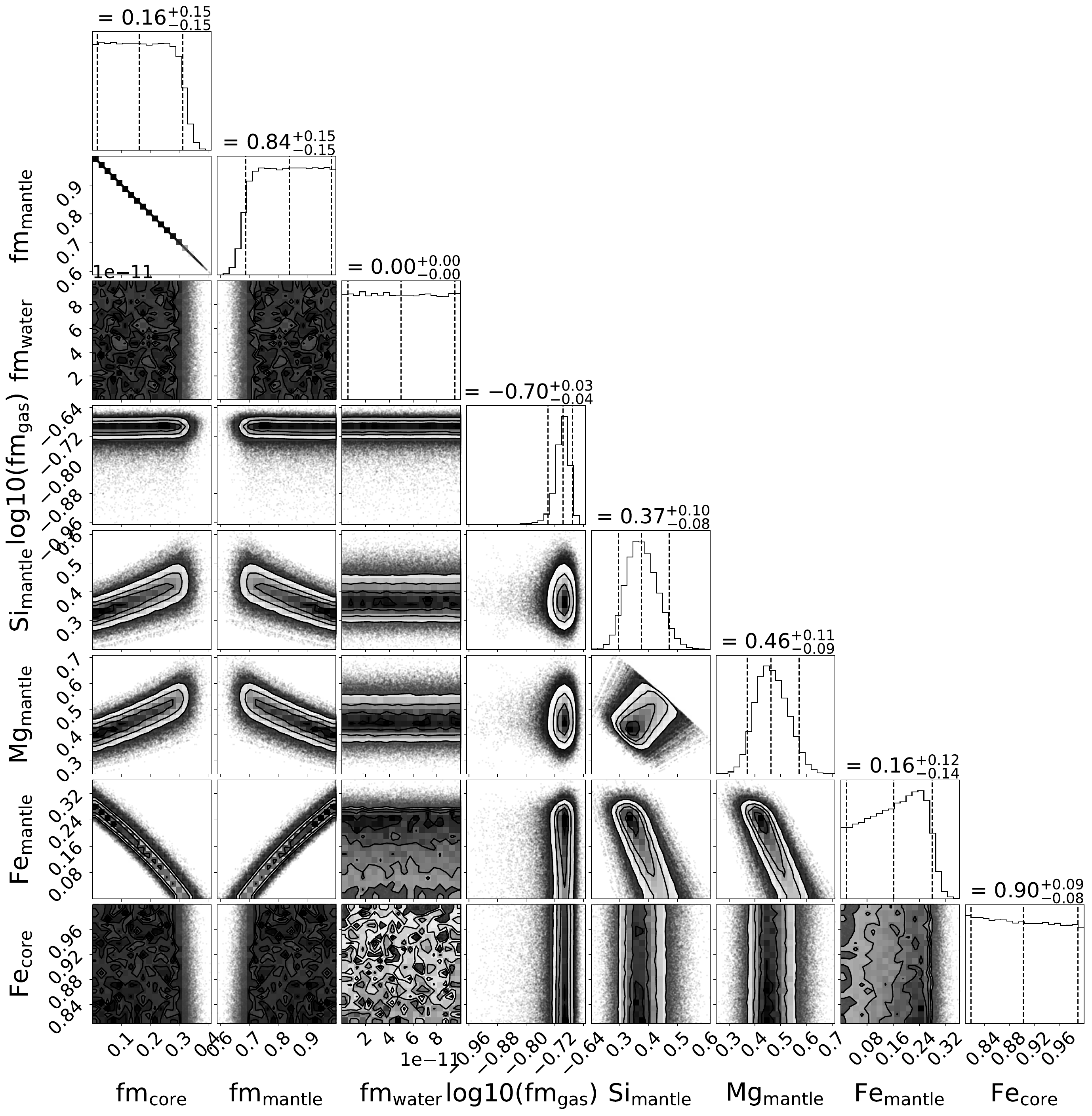}
\centering
\end{minipage}
\caption[]{Correlation diagram for the PDFs of the \textsc{biceps} parameters for the interior model of TOI-421~b (dry case).}
\label{fig:BICEPS_int_c_dry}
\end{figure*}


\section{Correlation diagrams for the RMR analysis}
\label{apn:corr_cont}

\begin{figure*}
\begin{minipage}[h!]{\textwidth}
\includegraphics[trim=0cm 0cm 0cm 0cm,clip=true,width=\columnwidth]{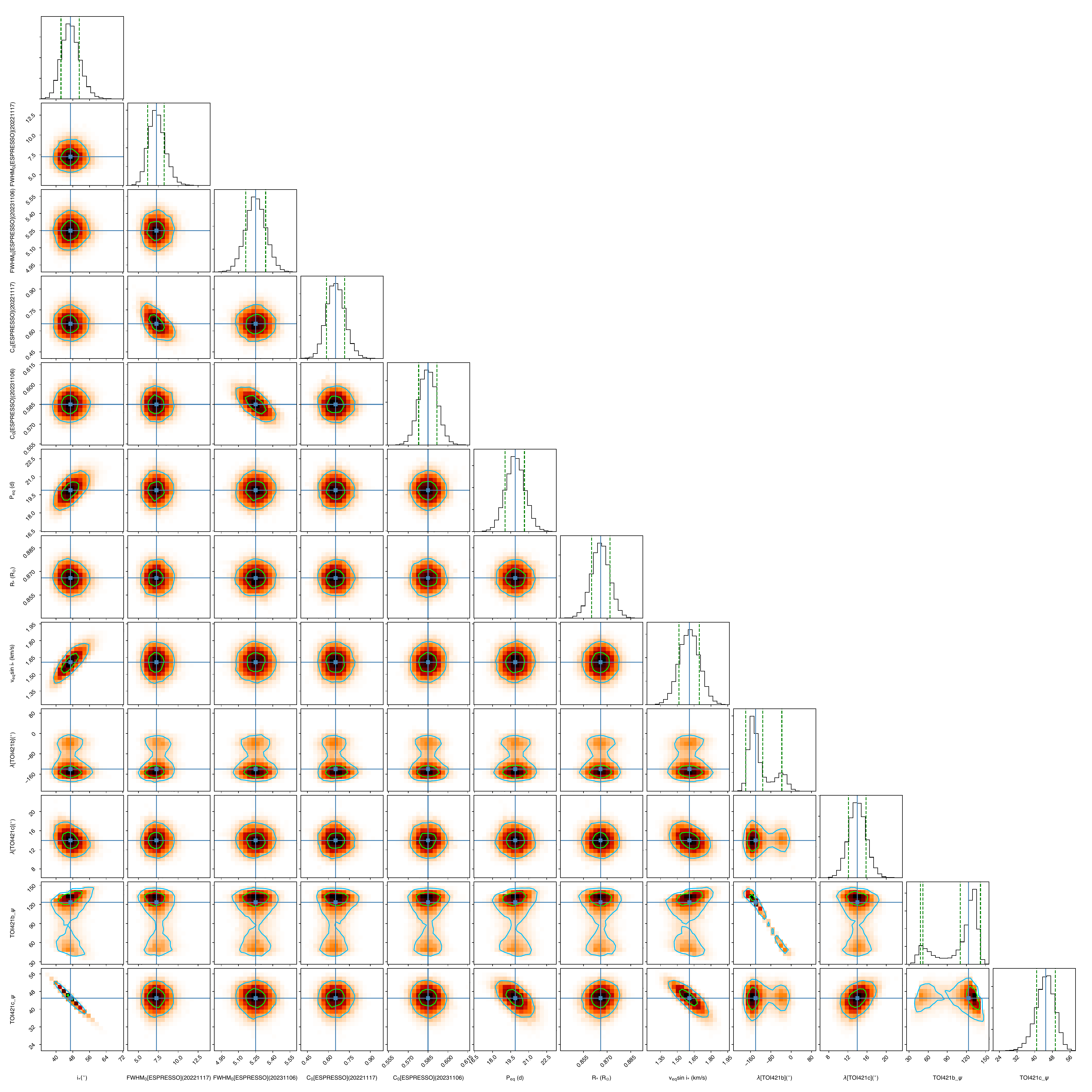}
\centering
\end{minipage}
\caption[]{Correlation diagrams for the PDFs of the RMR model parameters of the joint TOI-421~b and c transits, with $P_\mathrm{eq}$ as jump parameter. We note that the PDFs for $\lambda$, $v_\mathrm{eq} \sin i_{\star}$, and the line shape properties remain the same as in the nominal case. For completeness we show here both modes for $\lambda_\mathrm{b}$. Green and blue lines show the 1 and 2$\sigma$ simultaneous 2D confidence regions that contain, respectively, 39.3\% and 86.5\% of the accepted steps. 1D histograms correspond to the distributions projected on the space of each line parameter, with the green dashed lines limiting the 68.3\% HDIs. The blue lines and squares show the median values.}
\label{fig:Corr_diag_TOI421}
\end{figure*}

\end{appendix}

\end{document}